\documentclass[namedreferences]{solarphysics}

\usepackage[hyperref,optionalrh,showbiblabels]{spr-sola-addons} 
\usepackage{graphicx}        
\usepackage{color}           
\usepackage{breakurl}        
\usepackage{rotating}
\usepackage{lscape}
\usepackage{enumerate}

\newcommand{\etal}{{\it et al.}}

\usepackage{multirow}

\hypersetup{pdftitle = The title of my PDF, pdfauthor = My name, pdfsubject= The subject, pdfkeywords = keyword1 keyword2 keyword3} 
\hypersetup{colorlinks = true, linkcolor = red, anchorcolor = red, citecolor = blue, filecolor = red, pagecolor = red, urlcolor = red}

\newcommand{\aap}{    {\it Astron. Astrophys.}}

\newcommand{\apj}{    {\it Astrophys. J.}}
\newcommand{\apjl}{   {\it Astrophys. J. Lett.}}

\newcommand{\solphys}{{\it Solar Phys.}}

\chardef\us=`\_

\begin{document}

\begin{article}
\begin{opening}

\title{Onboard Automated CME Detection Algorithm for Visible Emission Line Coronagraph on ADITYA-L1\\ {\it Solar Physics}}

\author[addressref={aff1},corref,email={ritesh.patel@iiap.res.in}]{\inits{R.}\fnm{Ritesh}~\lnm{Patel} \orcid{0000-0001-8504-2725}}
\author[addressref=aff2,email={jrf0019@isac.gov.in}]{\inits{Amareswari}\fnm{Amareswari}~\lnm{K.} \orcid{0000-0002-7438-6212}}
\author[addressref={aff1,aff4},corref,email={vaibhav@iiap.res.in}]{\inits{V.}\fnm{Vaibhav}~\lnm{Pant} \orcid{0000-0002-6954-2276}}
\author[addressref={aff1,aff3},corref]{\inits{D.}\fnm{Dipankar}~\lnm{Banerjee} \orcid{0000-0003-4653-6823}}
\author[addressref={aff1,aff2,aff3},corref]{\inits{Sankarasubramanian}\fnm{Sankarasubramanian}~\lnm{K.}}
\author[addressref={aff1},corref]{\inits{Amit}\fnm{Amit}~\lnm{Kumar}}


\address[id=aff1]{Indian Institute of Astrophysics, Koramangala, Bangalore - 560034, India}
\address[id=aff2]{ISRO Satellite Centre, Bangalore - 560017, India}
\address[id=aff4]{Center for mathematical Plasma Astrophysics, KU Leuven, 3001, Belgium}
\address[id=aff3]{Center of Excellence in Space Sciences, IISER Kolkata - 741246, India}

\runningauthor{{ Patel R. \etal}}
\runningtitle{{ Automated Onboard CME Detection}}

\begin{abstract}
ADITYA-L1 is India's first space mission to study the Sun from Lagrangian 1 position. { \textit{Visible Emission Line Coronagraph}} (VELC) is one of the seven payloads in ADITYA-L1 mission scheduled to be launched around 2020. One of the primary objectives of the VELC is to study the dynamics of coronal mass ejections (CMEs) in the inner corona. This will be accomplished by taking high resolution ($\approx$ 2.51 arcsec pixel$^{-1}$) images of corona from 1.05 R$_{\odot}$ -- 3 R$_{\odot}$ at high cadence of 1 s in 10 \AA\ passband centered at 5000 \AA. Due to limited telemetry at Lagrangian 1 position we plan to  implement  an onboard automated CME detection algorithm. The detection algorithm is based on the intensity thresholding followed by the area thresholding in successive difference images spatially re-binned to improve signal to noise { ratio}. We present the results of the application of this algorithm on the data from existing coronagraphs such as { STEREO/SECCHI} COR-1, { which is space based coronagraph,} and K-Cor, { a ground based coronagraph,} because they have field of view (FOV) nearest to VELC. Since, no existing space-based coronagraph has FOV similar to VELC, we have created synthetic coronal images for VELC FOV after including  photon noise and injected CMEs of different types. The performance of CME detection algorithm is tested on these images. { We found that for VELC images, the telemetry can be reduced by a factor of 85\% or more keeping CME detection rate of 70\% or above at the same time.} Finally, we discuss the advantages and disadvantages of this algorithm. The application of such onboard algorithm in future will enable us to take higher resolution images with improved cadence from space and also reduce the load on limited telemetry at the same time. This will help in better understanding of CMEs by studying their characteristics with improved spatial and temporal resolutions.

\end{abstract}
\end{opening}

\section{Introduction}
     \label{S-Introduction} 
{ Any observable difference in coronal structure with time scale varying from few minutes to several hours, associated with a new, discrete bright feature propagating outwards in coronagraph field of view (FOV) is called as coronal mass ejection (CME)} (\citealp{Hundhausen84}).
The frequency of occurrence of these events vary from one per day during solar minimum to several times a day during solar maximum. 
The first observation of CMEs from space was in 1971 (\citealp{Hansen71}) by a white light coronagraph having FOV from 3 R$_{\odot}$ to 10 R$_{\odot}$ onboard { \textit{Orbiting Solar Observatory-7}} (\citealp{OSO75}). Later several other coronagraphs were sent to space. First statistical survey of CMEs was done with data obtained from coronagraph onboard { \textit{Skylab}} (\citealp{SKYLAB}) having FOV from 1.5~R$_{\odot}$ to 6 R$_{\odot}$. White light coronagraph with FOV from 2.6 R$_{\odot}$ to 10 R$_{\odot}$ onboard { \textit{Solwind}} (\citealp{SolWind}) recorded halo CME for the first time. A white light coronagraph having FOV from 1.5 R$_{\odot}$ to 6 R$_{\odot}$ and spatial resolution 10 arcsec, onboard { \textit{Solar Maximum Mission}} (\citealp{SMM}) studied over 1000 CMEs in its lifetime. The classic three-part CME was also identified during its course of operation. Coronagraph onboard { \textit{Spartan 201}} (\citealp{Spartan}) with FOV 1.25 R$_{\odot}$ to 6 R$_{\odot}$ flew five times between April 1993 and November 1998 and maintained the continuity of coronal data. { \textit{Large Angle Spectroscopic COronagraph}} (LASCO) C2 and C3 onboard { \textit{Solar and Heliospheric Observatory}} (SOHO) (\citealp{Brueckner95}) launched in 1995 has provided data continuously for over two solar cycles with FOV from 2 R$_{\odot}$ to 30 R$_{\odot}$. Most of our understanding about CMEs and variation in their characteristics over different phase of solar cycle has come from the analysis of data obtained from LASCO over the years. { \textit{Sun Earth Connection Coronal and Heliospheric Investigation}} (SECCHI) (\citealp{Howard02}) onboard { \textit{Solar Terrestrial Relations Observatory}} (STEREO) are pair of white light coronagraphs having FOV 1.5~R$_{\odot}$ to 16 R$_{\odot}$  provide coronal observations from two different vantage points. The only coronagraph to have FOV closest to solar limb was LASCO C1 (1.1~R$_{\odot}$ to 3 R$_{\odot}$) which stopped its operation two years afters its launch in 1998. Though the launch of these coronagraphs in space have increased our knowledge of solar corona and coronal transients, the inner corona is not explored in great details.

{ \textit{Visible Emission Line Coronagraph}} (VELC) (\citealp{Singh2011,Singh13,VELC17}) onboard { \textit{ADITYA-L1}} (\citealp{ADITYA2017}) has FOV ranging from 1.05 R$_{\odot}$ to 3 R$_{\odot}$. It will provide an unique opportunity to probe the inner corona. VELC will take images of corona with high resolution of 2.51 arcsec pixel$^{-1}$ at wavelength of 5000 \AA\ with cadence that can go up to 1 s. This combined with spectroscopy and spectro-polarimetric capabilities will allow us to understand the origin of CMEs and their behaviour in the inner corona.

CMEs have been visually inspected and cataloged with their measured properties (apparent central position angle, average angular width, height as function of time) using the continuous data obtained from LASCO coronagraph. One of the earliest CME catalog was made by \cite{Yashiro04} and \cite{Gopalswamy09} based on visual inspection of LASCO data and is available online\footnote{\url{http://cdaw.gsfc.nasa.gov/CME_list}}. Since last few years the amount of data generated from space-based coronagraphs has increased tremendously. Moreover, visual inspection process is labour intensive and subjective due to which several onground automated CMEs detection algorithms were developed. Computer Aided CME Tracking (CACTus) was the first of such techniques that use Hough transform to detect CMEs in polar transformed running difference coronagraph images (\citealp{Robbrecht04}, \citealp{2016ApJ...833...80P}). Initially it was applied in LASCO C2 and C3 images, but later its application was extended to STEREO COR-2 and { \textit{Heliospheric Imager-1}} images also. Solar Eruptive Events Detection System (SEEDS) developed by \cite{Olmedo08}, automatically detects CMEs in polar transformed running difference images of LASCO and COR-2 with threshold-segmentation technique to extract an appropriate shape of the leading edge of CMEs. Another such algorithm developed for automated CMEs detection is Automatic Recognition of Transient Events and Marseille Inventory from Synoptic maps (ARTEMIS) (\citealp{ARTEMIS}) that detects CMEs automatically in synoptic maps based on adaptive filtering and segmentation. COronal IMage Processing (CORIMP) algorithm for automated CMEs detection is based on separation of quiescent and dynamic structures in coronagraph images using deconvolution followed by multi-scale edge detection method to detect and track CMEs (\citealp{Morgan12,Byrne12}). CORIMP has been successfully implemented for automated CMEs detection in LASCO and COR-2 images. { \cite{Texture2010} used the texture of CMEs in white light images for CME detection and tracking. Machine learning algorithms have also been used to detect and track CMEs based on multiple features (\citealp{MLQu2006, MLADABOOST16, ML2017}).CMEs in heliosphere are identified manually as well as automatically in STEREO/HI-1 images and have been catalogued as a part of Heliospheric Cataloging Analysis and Techniques Service (HELCATS)\footnote{\url{https://www.helcats-fp7.eu/}} project (\citealp{Harrison2018}). The first multi-viewpoint CME catalog\footnote{\url{http://solar.jhuapl.edu/Data-Products/COR-CME-Catalog.php}} by visually inspecting SECCHI/COR-2 total brightness images has been made by \cite{COR2CME}.}

With the advancement of technology, the quality of high resolution imaging with high cadence has been achieved. However, due to the limitation in the download bandwidth from deep space, the telemetry load needs to be reduced. This can be overcome by on-board detection of CMEs and sending images that contain only CMEs. Though these on-ground automated CMEs detection methods are very robust, but they are too memory intensive to be applied in onboard electronics. Among the upcoming solar missions, an automated CMEs detection has been developed by \cite{Bemporad14} for  CMEs detection in { \textit{Multi Element Telescope for Imaging and Spectroscopy}} (METIS) onboard { \textit{Solar Orbiter}} (\citealp{METIS12}). This algorithm is based on running difference between consecutive images re-binned to low resolution followed by intensity thresholding. It takes the advantage of white light coronal images obtained at different polarization angles so that even a small variations in white light intensity during CMEs can be detected.

Since VELC is designed to take high resolution ($\approx$ 2.51 arcsec pixel$^{-1}$) images of corona with 1 s cadence, the amount of data generated will exceed 1 TB per day for the continuum channel alone. Therefore, a simple algorithm has been developed to reduce the telemetry load. The algorithm is based on intensity threshold in re-binned running difference coronagraph images to bring out the dynamic coronal features such as CMEs, from the background followed by area threholding in order to capture CMEs and discard any artifact which could have been detected as CMEs by intensity thresholding.

In this paper we first describe the synthesis of background corona for VELC field of view in Section \ref{S-synthetic}. The expected images for VELC after inclusion of noise and scattered intensity along with addition of CMEs of different types are also presented in this section. The onboard detection algorithm is explained in Section \ref{S-Algo}. Its application on data of existing coronagraphs, STEREO COR-1A and K-Cor, is illustrated in Section \ref{S-Results existing} while the application on simulated data is presented in Section \ref{S-Results simulated}. Finally the conclusions are drawn  in Section \ref{S-Conclusion}.

\section{Synthetic Corona and CMEs}
     \label{S-synthetic}
     
VELC has FOV extending from 1.05 R$_{\odot}$ to 3 R$_{\odot}$. No other existing space-based coronagraphs take images of corona in white light at this height.{ \textit{Association de Satellites pour l’Imagerie et l’Interferométrie de la Couronne Solaire} (ASPIICS) onboard \textit{PROBA-3} (\citealp{Proba3}) having FOV from 1.08$_{\odot}$ to 3$_{\odot}$ similar to VELC might be launched around the same time of VELC.} One coronagraph, K-Cor has FOV similar to VELC. However, the images obtained from K-Cor have large contributions from the scattered intensity due to atmosphere. Though we have tested our algorithm on these images, but it suffer from false detection and has been discussed in section \ref{S-kcor}. This led to the need for creating synthetic images of corona for VELC FOV and adding CMEs of different properties for the purpose of testing the performance of onboard automated CME detection algorithm. 

\subsection{Preparation of Synthetic Coronal Images} 
  \label{S-prepcor}
The Hulst model of solar corona (\citealp{Hulst50}) for minimum phase of solar cycle is used to create synthetic coronal images as VELC will observe corona during solar minima. This model separates the coronal intensities in equatorial and polar regions with radial variations as given by Equations \ref{Hulst_eq} and  \ref{Hulst_pl} respectively.
\begin{equation} 
\label{Hulst_eq}
B(x) = (200.00x^{-17} + 114.86x^{-7} + 5.38x^{-2.5})\times 10^{-8} \mathrm{{B_\odot}},
\end{equation}
\begin{equation} 
\label{Hulst_pl}
B(x) = (191.00x^{-17} + 27.45x^{-7} + 4.99x^{-2.5})\times 10^{-8} \mathrm{{B_\odot}},
\end{equation}
where, \textit{x} is the distance in units of solar radius and { B\textsubscript{$\odot$}} is mean solar disk brightness.

The size of the image is taken as $2160\times2560$ which is same as the size of detector to be used in VELC. The pixel scale of VELC continuum channel is 2.51 arcsec pixel$^{-1}$. Since, the detector is rectangular in shape, the shorter side will cover 1080 pixel radius equivalent to 2.81 R$_{\odot}$. The images are simulated for $2560\times2560$ and cropped based on the orientation of the detector to choose the shorter side along equator or pole.

It has been reported by \cite{Hulst50} that mostly K corona dominates up to 2 R$_{\odot}$. It has also been observed that during the period of maximum activity of solar cycle, the white light corona (K+F) is almost uniformly distributed around the solar limb while during minimum activity period, it is mostly confined to the equatorial regions and hence the shape looks nearly elliptical in white light images (\citealp{Marzouk16}). In order to incorporate this shape, a synthetic coronal image is first made using Equation \ref{Hulst_eq} with circularly symmetric intensity (B). Next B is modulated with latitude variation in coronal intensity as given by Equation \ref{modHulst}, 
\vspace{-0.10cm}
\begin{equation} 
\label{modHulst}
B_1(q,z) = \frac{5\cos{\frac{\theta}{2}}}{\sin{\frac{\theta}{2}}+0.5} + \mathrm{{exp}}(-\pi \sin^{2}\theta) + \frac{1}{8 \pi \sigma^{2}} \mathrm{{exp}}(\frac{-\pi \sin^{2}\theta}{4k\sigma^{2}})\times B(q,z),
\end{equation}
where, $1 \leq q,z \leq 2560$ and $\theta= \tan^{-1}(\frac{q-cx}{z-cy})$. $\theta$ is the angle with respect to horizontal axis.

The motivation of using this relation is to get a nearly elliptical shape of white light corona in synthetic images. The first and second term in Equation \ref{modHulst} governs the extent and spread of increased brightness at the equator, the third term adds the gradual variation in the brightness while moving from the equator to poles. Here, $\sigma$ combined with k defines the sharpness in the asymmetry. The shape and the amount of asymmetry can be changed by changing the values of $\sigma$ and k. Very low values of $\sigma$ and k leads to very sharp intensity increase in the equatorial region. We note that increasing the values of $\sigma$ and k, increases the circular symmetry of the intensity of the corona. After some trials, $\sigma$ and k are chosen as 0.15 and 12 respectively for the closest match to the observed coronal intensities.
B(q,z) is the intensity value at [q,z] pixel position of the circular symmetric image, B$_1$(q,z) is the modified intensity after introducing asymmetry. The image B$_1$ is then combined with the original image B using Equation \ref{BG} such that the resulting image has the gradual variation in intensity from equator to poles.
\vspace{-0.10cm}
\begin{equation} \label{BG}
BG= \frac{1}{4}[ 2.05B + 0.05B_1 ],
\end{equation}
where BG is the final synthetic coronal image for VELC FOV. The coefficients of B and B$_1$ are obtained after few trials in order to get the intensity values closer to the observed values. 

\begin{figure}[!ht]
\vspace{-0.05\textwidth}
   \centerline{\includegraphics[width=0.5\textwidth,clip=]{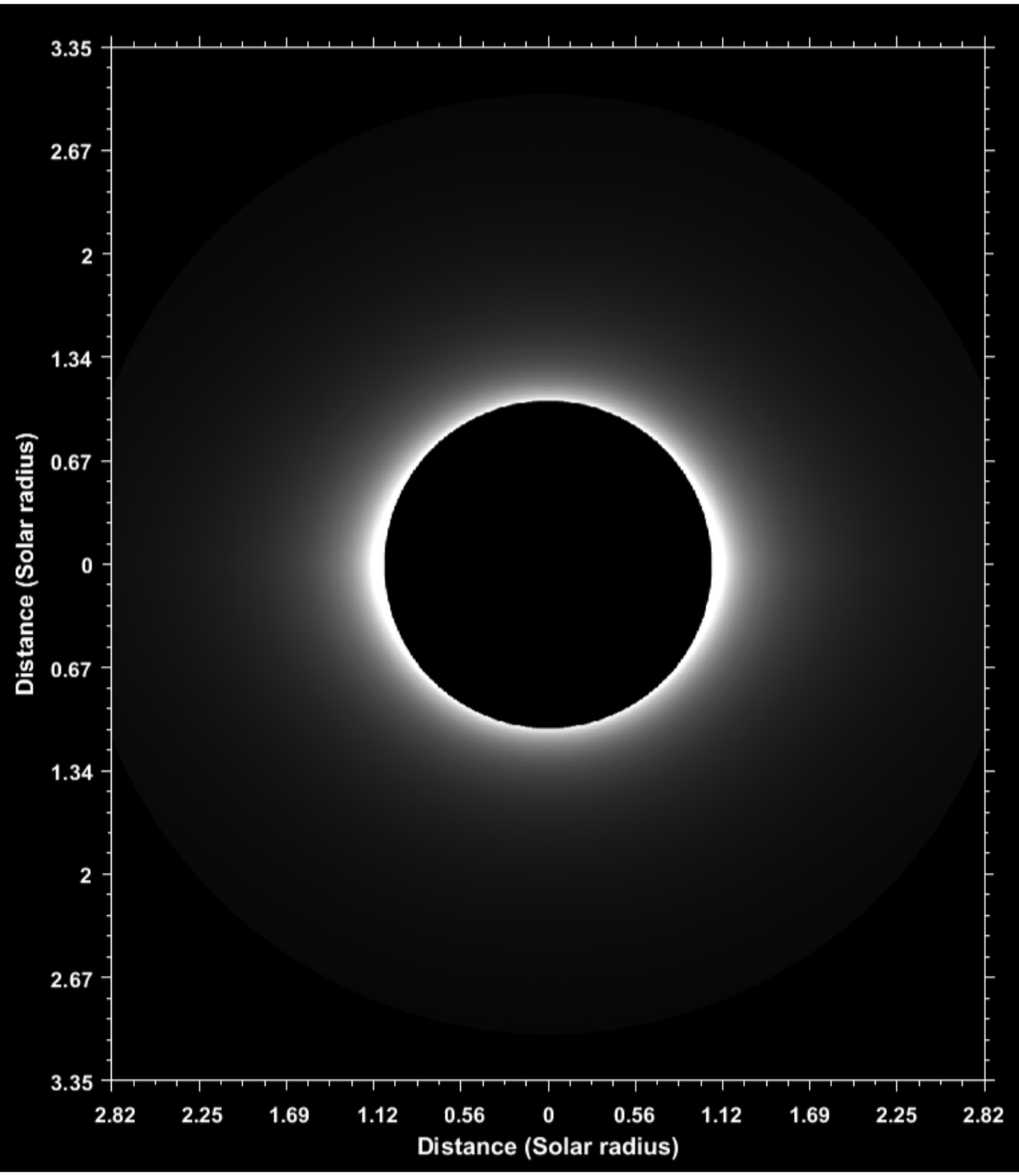}
              }
              \vspace{-0.05\textwidth}
              \caption{Synthetic coronal image for VELC FOV taking $\sigma$=0.15 and k=12. The axes are represented in solar radii. Equator is along the { horizontal} axis.
                      }
   \label{VELCfov}
   \end{figure}
   
Figure \ref{VELCfov} shows a synthetic coronal image for VELC FOV taking $\sigma$=0.15 and k=12. It should be noted that the image is of size $2160 \times 2560$ and the equator is along the { horizontal} axis.

\begin{figure}[!h] 
\centerline{\hspace*{0.015\textwidth}
               \includegraphics[width=0.51\textwidth,clip=]{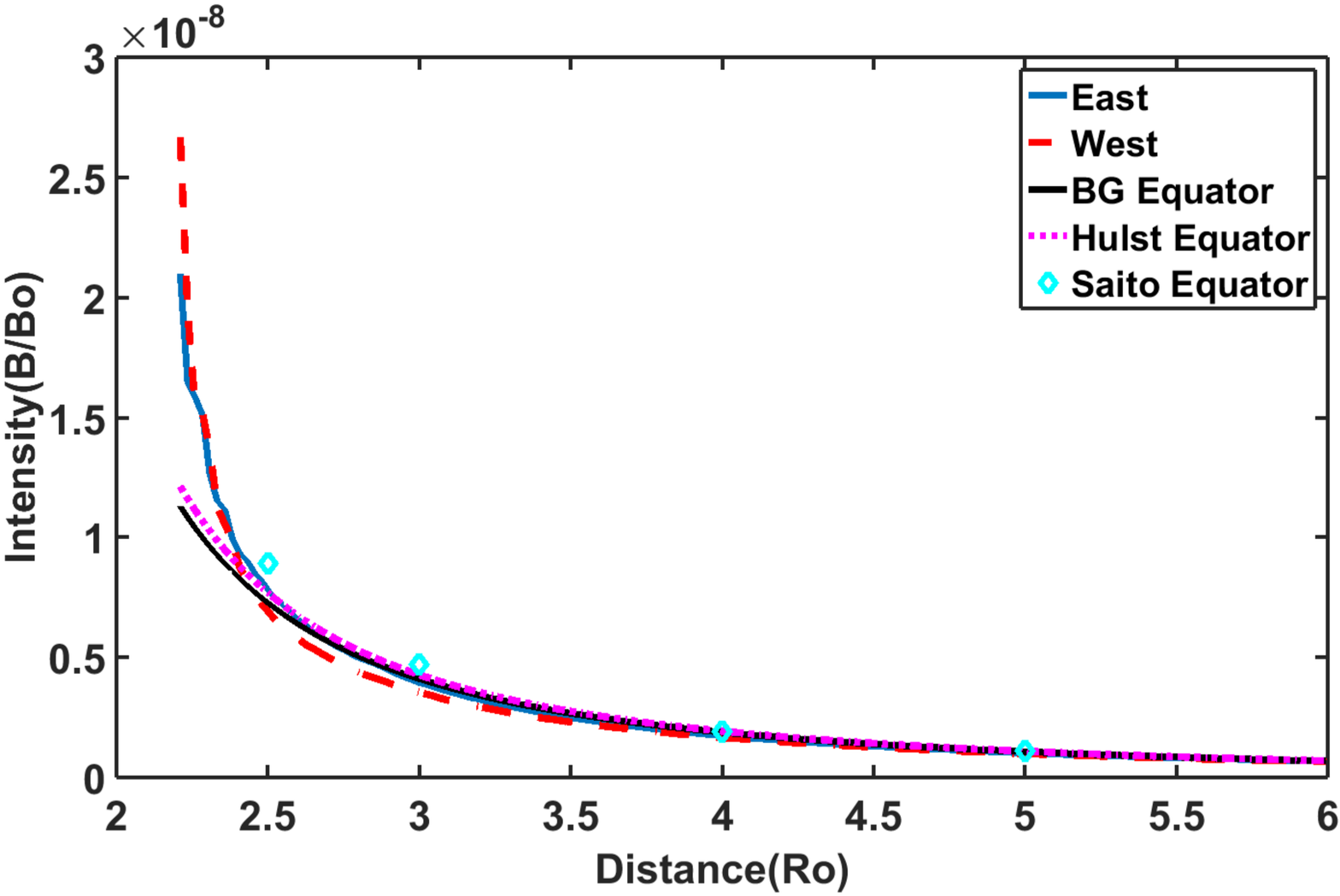}
               \hspace*{-0.015\textwidth}
               \includegraphics[width=0.51\textwidth,clip=]{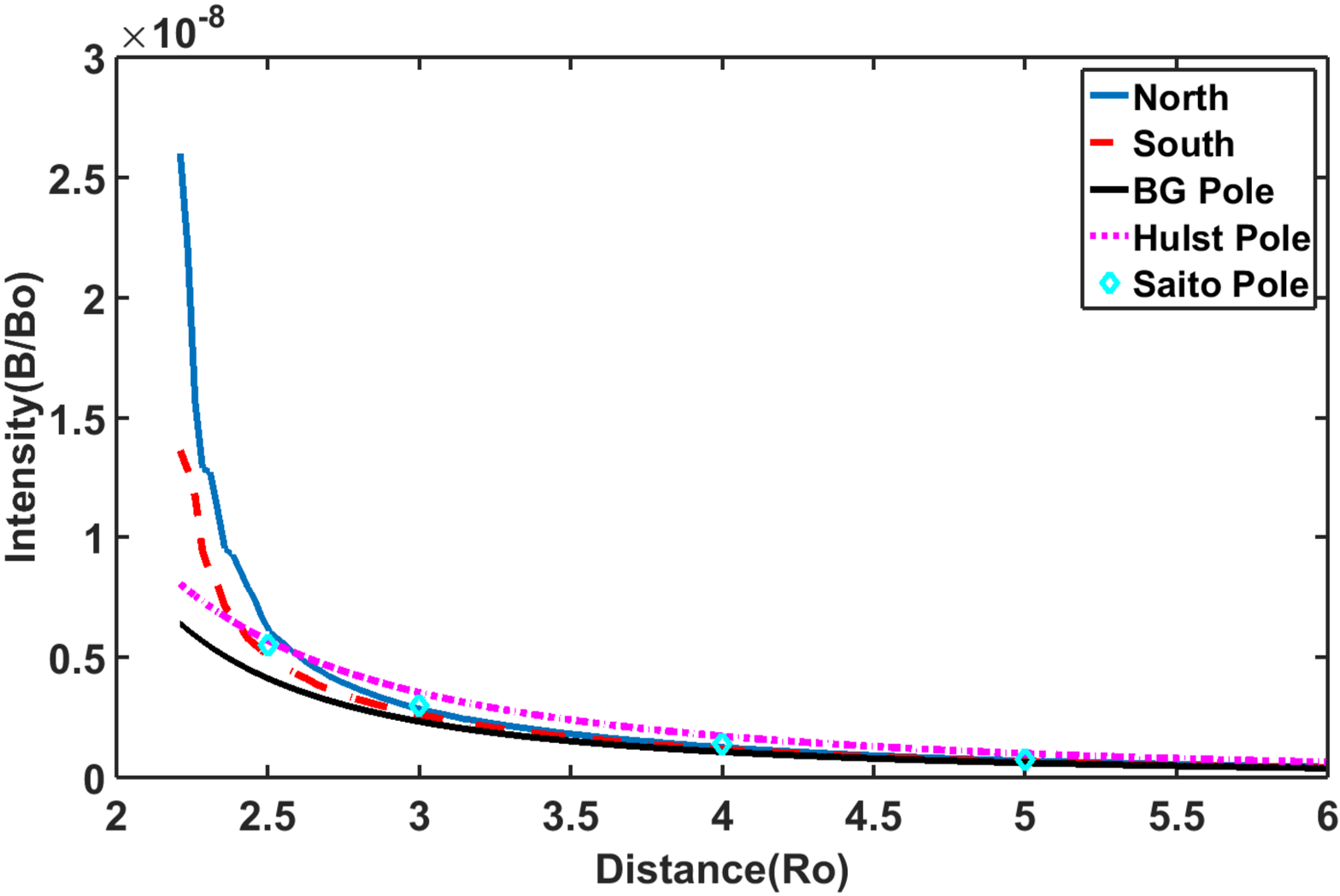}
              }
     \vspace{-0.005\textwidth} 
     \caption{Comparison of coronal intensities calculated by our modified Hulst model represented by BG with LASCO C2 image taken on 01-01-2009 and theoretical corona model intensities given by Hulst and Saito. $\it{Left}$ Coronal intensities at equator. $\it{Right}$ Coronal intensities at poles. The solid black colored line corresponds to corona intensity with modified Hulst model, solid line in blue and dashed line in brown represents intensities for LASCO C2, dotted line represents coronal intensities given by Hulst while those given by Saito are marked by plus.}
     \label{comparison}
\end{figure}

{ As VELC will observe the solar corona during rising phase of solar cycle, we compared our corona model with the observational data of LASCO C2 during this phase.} Figure \ref{comparison} shows the comparison of intensities of the synthetic corona image prepared by our corona model, with the intensity observed by LASCO C2 and theoretical models given by Hulst and Saito. The polarised brightness images of 01-01-2009 captured by LASCO C2 was taken to create total brightness image. Average intensities along the radial direction at equator and poles were obtained by taking the mean intensity of three pixels width. This was compared with the average intensity at equator and pole of the synthetic image simulated by our model, with average intensity calculated in same way like LASCO C2. The coronal intensity values given by theoretical models of Hulst and Saito are also compared for equatorial and polar region intensities. It can be seen that the intensities obtained by our model for synthetic image match well with the existing theoretical models as well as with the observed ones. The deviation of the curves near the inner edge of FOV may be accounted for the high scattered intensities and diffraction pattern in this region of image which contaminates the corona signal.

\begin{figure}[!ht]
\vspace{-0.2\textwidth}
  \centerline{\includegraphics[width=1\textwidth,clip=]{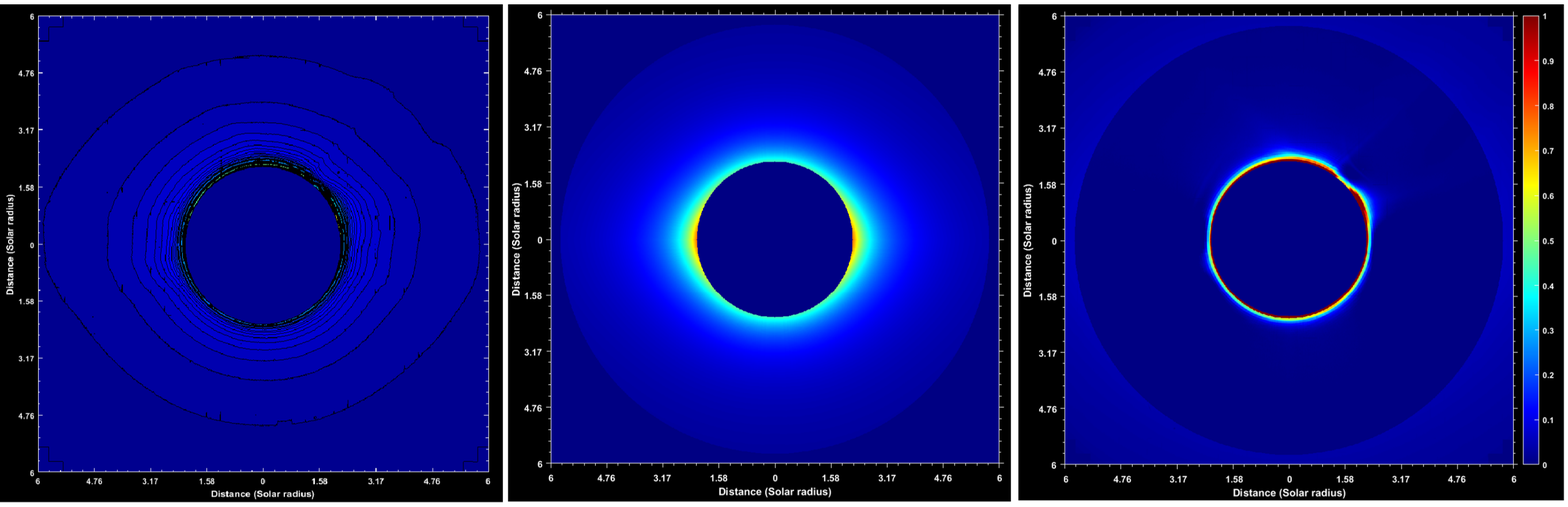}
              }
              \vspace{-0.175\textwidth}
              \caption{\textit{Left:} Intensity contour of 11 days minimum background image obtained from LASCO C2 images taken from 01/01/2009 to 11/01/2009. \textit{Centre:} Synthetic corona image generated for C2 FOV using our model. \textit{Right:} Residual intensity obtained from subtracting the synthetic coronal image generated for C2 FOV and 11 days minimum background image.
                      }
  \label{C2diff}
  \end{figure}

{ The  minimum background corona is calculated using standard technique in which minimum image is prepared from 11 days coronagraph images. This is used to remove some static and semi-static coronal features like streamers which die out during this time giving approximately a coronal background (\citealp{DeForest11, DeForest14}).} Left panel in Figure \ref{C2diff} shows an image of the intensity contour of 11 days minimum background image obtained from C2 images taken from 01/01/2009 to 11/01/2009 followed by synthetic coronal image generated for C2 FOV in centre panel. The difference image of the two is shown in the right panel. It can be seen that except for the bright coronal features (due to streamers), the difference in the intensities from the observation data and synthetic corona model is minimal. The mean intensity of the residual image is found in the order of 10$^{-10}$ B$_{\odot}$. This further validates our model to be used for generation of synthetic coronal images for VELC FOV.

The synthetic coronal images so generated have intensity values in units of brightness (W m$^{-2}$ sr$^{-1}$ nm$^{-1}$). However, the data recorded onboard will be in terms of data numbers (DN). So, the intensity is converted to photoelectrons thereby converting it to DN before applying the CME detection algorithm. The conversion is done by using the Planck$'$s black body radiation relation to find the mean brightness recorded in each pixel and then converting to DN by taking 15 electrons per DN for VELC CMOS detector. This relation is given by Equation~\ref{ne} assuming the efficiency of continuum channel as 25\% for 5000 \AA,

\vspace{-0.50cm}
\begin{equation} \label{ne}
    n = 3.863\times10^{10} I, 
\end{equation}
where $\textit{{ I}}$ is obtained from coronal intensity with radial variation curve and is in order of $10^{-6}$ with units of photoelectrons pixel$^{-1}$ s$^{-1}$.

 The coronagraph images have significant contribution from the instrument scattered light. Therefore, the scattered light contribution needs to be accounted for in the synthetic coronal images. The scatter studies for the VELC continuum channel has been done in Advanced System Analysis Program (ASAP), as reported in \citealp{Venkata17}. The data is in units of mean solar disk brightness with the radial variation within VELC field of view. A two dimensional image is made using this data assuming a circular symmetry. This image of scattered intensity is then added to the simulated coronal images. 
 
 \begin{figure}[!ht]
 \vspace{-0.03\textwidth}
   \centerline{\includegraphics[width=1\textwidth,clip=]{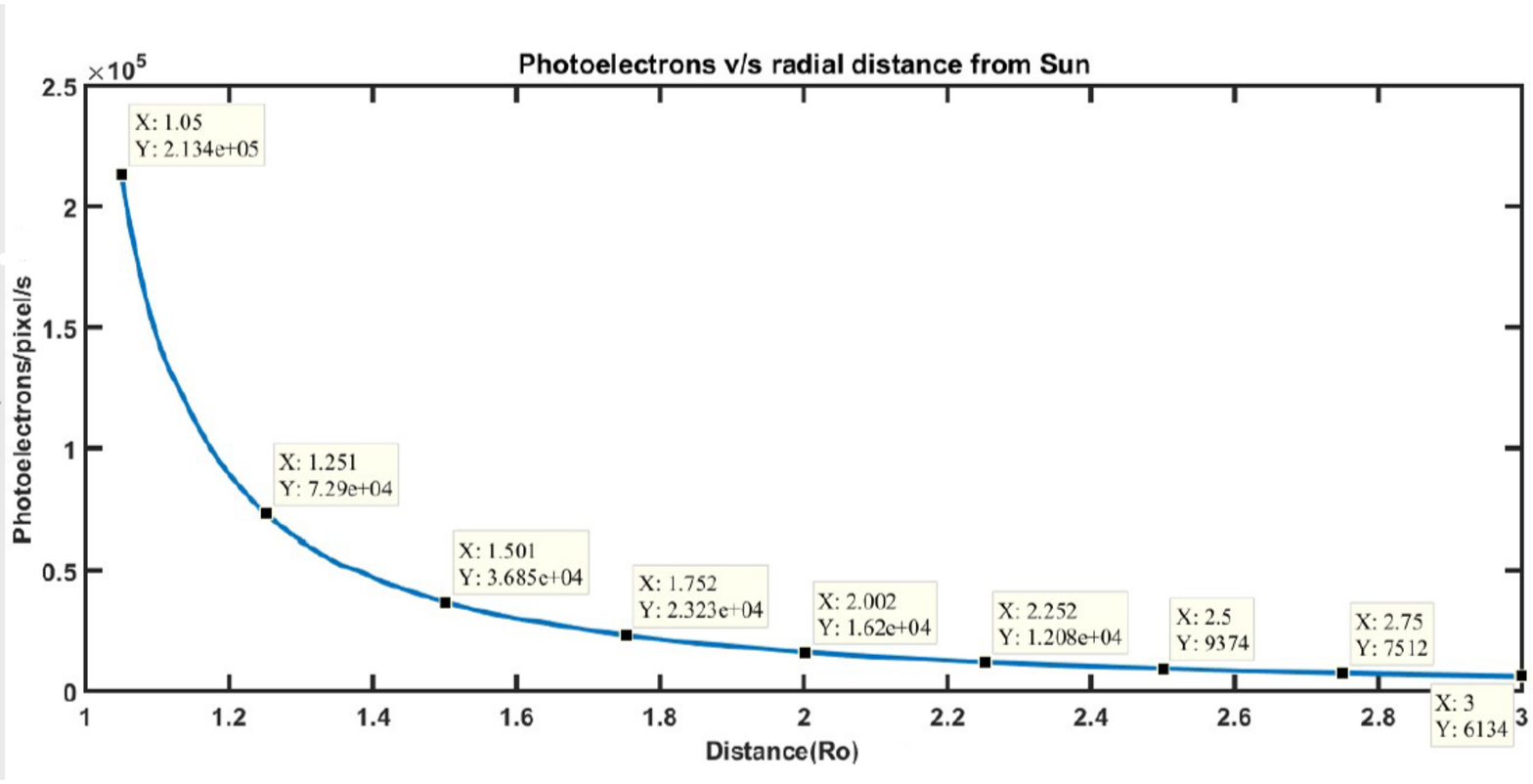}
              }
              \vspace{-0.1\textwidth}
              \caption{Variation of number of photoelectrons per pixel per second along radial direction in VELC FOV after addition of scattered intensity.
                      }
        \label{sctrad}
   \end{figure}
   
 Figure \ref{sctrad} shows the radial variation in the rate of generation of number of photoelectrons after incorporating the contribution of instrumental scattered intensity in the synthetic coronal image. It can be seen that at 1.05 R$_{\odot}$, where the intensity will be maximum, more than 213,400 photoelectrons will be generated every second. The full-well capacity of VELC continuum channel detector is 30000 electrons, which gives the exposure time of detector without saturation to be less than 140 ms. Therefore, the exposure time for this channel is set to be 100 ms. In order to build the coronal signal, subsequent frames will be added for further processing which is explained in section \ref{S-Algo}. The exposure time can be reduced in case of saturation of the detector if a bright CME passes through streamer.
  
The inclusion of scattered intensity is followed by the addition of noise in simulated images. Usually, the image sensor and the electronics related to it are the sources of noise. For the synthetic coronal images, two major sources of noise are identified as shot/photon noise and Gaussian noise (\citealp{McLean08}). Photon noise(p) due to the statistical fluctuation in the number of photons incident on the detector for a given exposure time is included as square root of the number of incident photons. Gaussian noise which is independent at each pixel, includes the noise due to thermal agitation of electrons even at room temperature also called dark noise (D) and readout noise (R). So, the total noise is given by Equation \ref{noise},
\vspace{-0.40cm}
\begin{equation} \label{noise}
(\mathrm{{noise}})^2 = p^2 + D^2 + R^{2}.
\end{equation}

\begin{figure}[!ht]   
   \centerline{\includegraphics[width=1\textwidth,clip=]{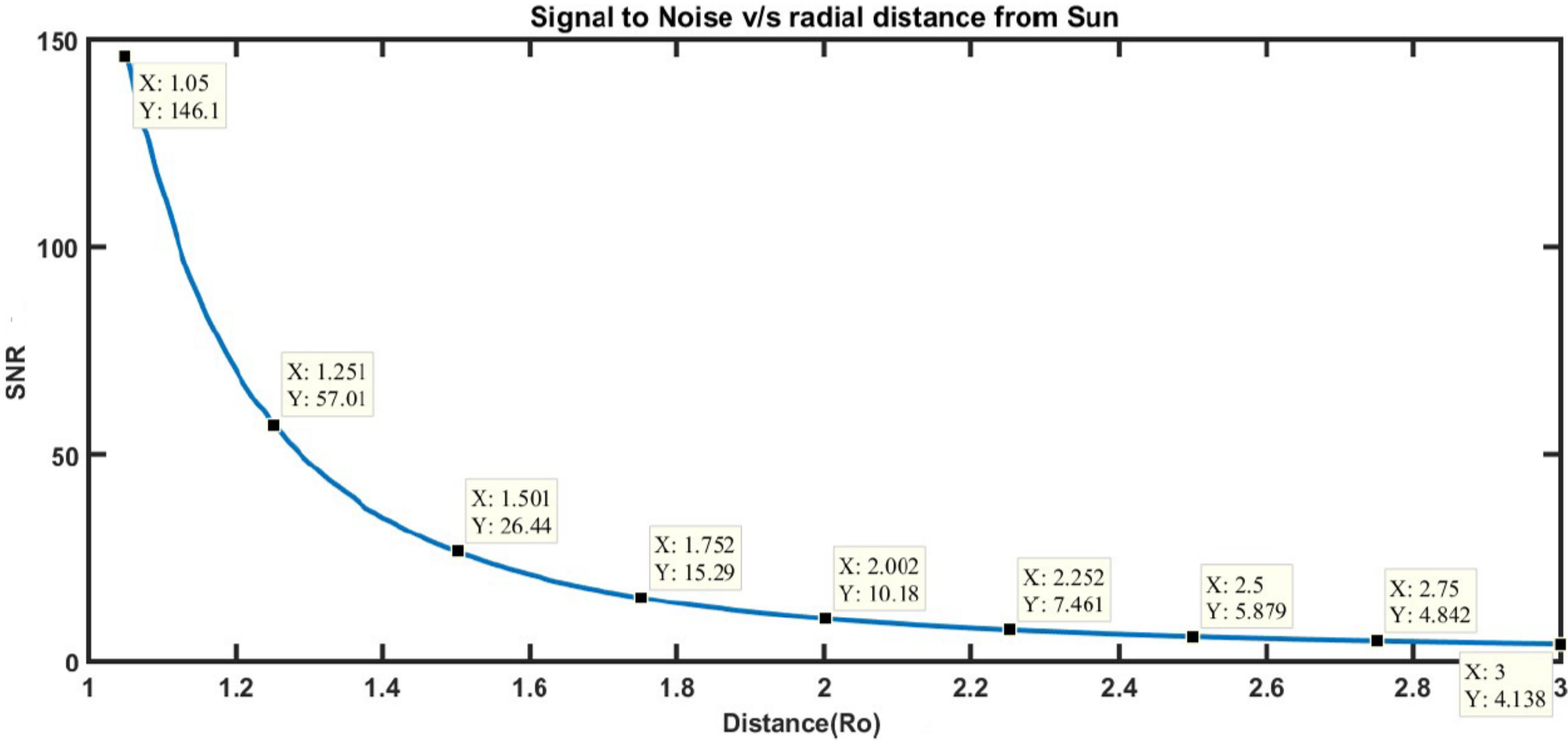}
              }
              \vspace{-0.1\textwidth}
              \caption{{Variation of signal to noise ratio with radial distance in VELC FOV.}
                      }
   \label{SNR_Plot}
   \end{figure}

The number of incident photons for photon noise is calculated by summing up the incident photons of the coronal brightness and the scattered intensity. The efficiency of the continuum channel including the optics and detector is 25\%. The RMS electrons for read out noise of VELC CMOS detector is 2 electrons where as dark noise will be less than 15 electrons. Combining all the noise, the signal to noise ratio (SNR) can be calculated as Equation \ref{SNR}, 
\vspace{-0.20cm}
\begin{equation} \label{SNR}
\mathrm{{SNR}} = \frac{S}{\sqrt{p^2 + D^2 + R^2}}.
\end{equation}
Where, S in the incident coronal signal. The variation of SNR with radial distance in VELC FOV is plotted as shown in Figure \ref{SNR_Plot}. It can be inferred from Figure \ref{SNR_Plot} that the signal to noise ratio in VELC reduces to 4 from 150 while moving from inner FOV to the edge of the image.
\vspace{-1.40cm}

The synthetic coronal images obtained after the inclusion of scattered intensity and noise will serve as the raw images over which CMEs of different types are added for our analysis.

\subsection{Simulation of CMEs in Coronal Images}
     \label{S-CME}
CMEs in coronagraph images appear as bright structure moving outwards. Thus we simulate CMEs in synthetic coronal images by injecting structure of brightness more than the background coronal intensity by few percent. It has been found that the brightness of CMEs with respect to the background varies from few percent to unity (\citealp{Robbrecht04}) which is defined as the intensity enhancement factor (xfac). The angle with respect to solar north at which CME has to be launched (phi) and the type of CME specified whether the CME is normal CME, narrow CME { or halo CME} are also defined {\it a priori}. 
{ Since, the average angular width of CMEs in inner corona is around 37$^{\circ}$ (\citealp{Cremades04}), so we simulated normal CMEs with width of 40$^{\circ}$ which expands as it propagates. We have also simulated symmetric halo CME with source region at the center of solar disk and asymmetric halo CME with source off center in the disk.}
The xfac is taken in the range of 0.05 to 0.75 to account for the faintest and the brightest CMEs in our simulation. The brightest CMEs can have xfac of unity, but we considered till 0.75 as CMEs brighter than this value are surely to be detected by our algorithm with the parameters discussed in Section \ref{S-Results simulated}. We have assumed the self similar expansion of CMEs and the leading edge of CMEs propagate depending on the ratio ($\kappa$) of the minor radius of flux rope to its major radius that remains approximately constant with time (\citealp{Subramanian14}). The value of $\kappa$ is varied from 0.2 to 0.4 to get different variants of CMEs. The average speed of the CMEs can be varied in our simulation by controlling the interval (N) between successive images. For an average speed  of $\approx$400 km s$^{-1}$, the value of N is set to be 1. We have generated CME images with average speed of $\approx$100 km s$^{-1}$, $\approx$400 km s$^{-1}$, $\approx$1600 km s$^{-1}$ and $\approx$2000 km s$^{-1}$ corresponding to slow, intermediate and fast speed CMEs respectively. { The  different  types of  CMEs  are simulated with  different  intensities,  angular  widths  and  average projected speeds and are listed out in Table \ref{CMEtypes}.}

\begin{table}[]
\caption{List of simulated CMEs}
\label{CMEtypes}
\begin{tabular}{llcccc}
\hline
\multicolumn{1}{c}{\begin{tabular}[c]{@{}c@{}}CME\\[1ex] No.\end{tabular}} & \multicolumn{1}{c}{\begin{tabular}[c]{@{}c@{}}CME \\[1ex] Type\end{tabular}} & \begin{tabular}[c]{@{}c@{}}Intensity \\[1ex] Enhancement \\[1ex] Factor\\[1ex] (xfac)\end{tabular} & \begin{tabular}[c]{@{}c@{}}Average\\[1ex] Speed\\[1ex] (km s$^{-1}$)\end{tabular} & {$\kappa$} & \begin{tabular}[c]{@{}c@{}}Position\\[1ex] Angle\\[1ex](Degree)\end{tabular} \\ \hline
CME 1 & Normal & 0.45 & 100 & 0.33 & 300 \\ \hline
CME 2 & Normal & 0.05 & 400 & 0.25 & 90 \\ \hline
CME 3 & Normal & 0.5 & 400 & 0.3 & 270 \\ \hline
CME 4 & Narrow & 0.05 & 400 & 0.23 & 80 \\ \hline
CME 5 & Narrow & 0.65 & 400 & 0.3 & 290 \\ \hline
CME 6 & Normal & 0.5 & 2000 & 0.38 & 300 \\ \hline
CME 7 & Normal & 0.75 & 2000 & 0.4 & 60 \\ \hline
{ CME 8} & { Halo (symmetric)} & { 0.5} & { 1600} & { 0.4} & { 360} \\ \hline
{ CME 9} & { Halo (asymmetric)} & { 0.6} & { 1600} & { 0.4} & { 60} \\ \hline
\end{tabular}
\end{table}

\begin{figure}[!ht]   
\vspace{-0.0225\textwidth} 
   \centerline{\hspace*{0.05\textwidth}
               \includegraphics[width=0.4\textwidth,clip=]{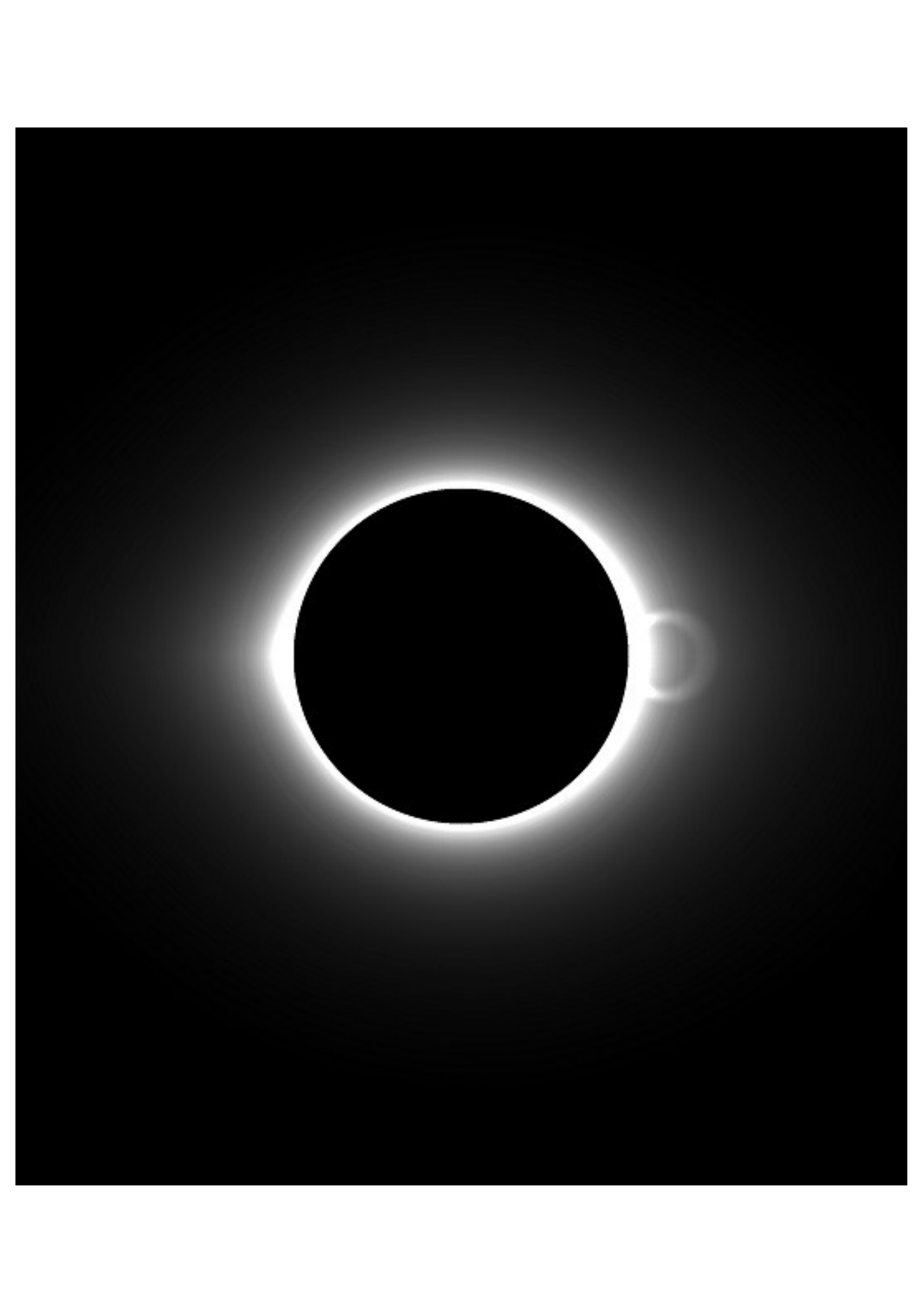}
               \hspace*{0.0001\textwidth}
               \includegraphics[width=0.4\textwidth,clip=]{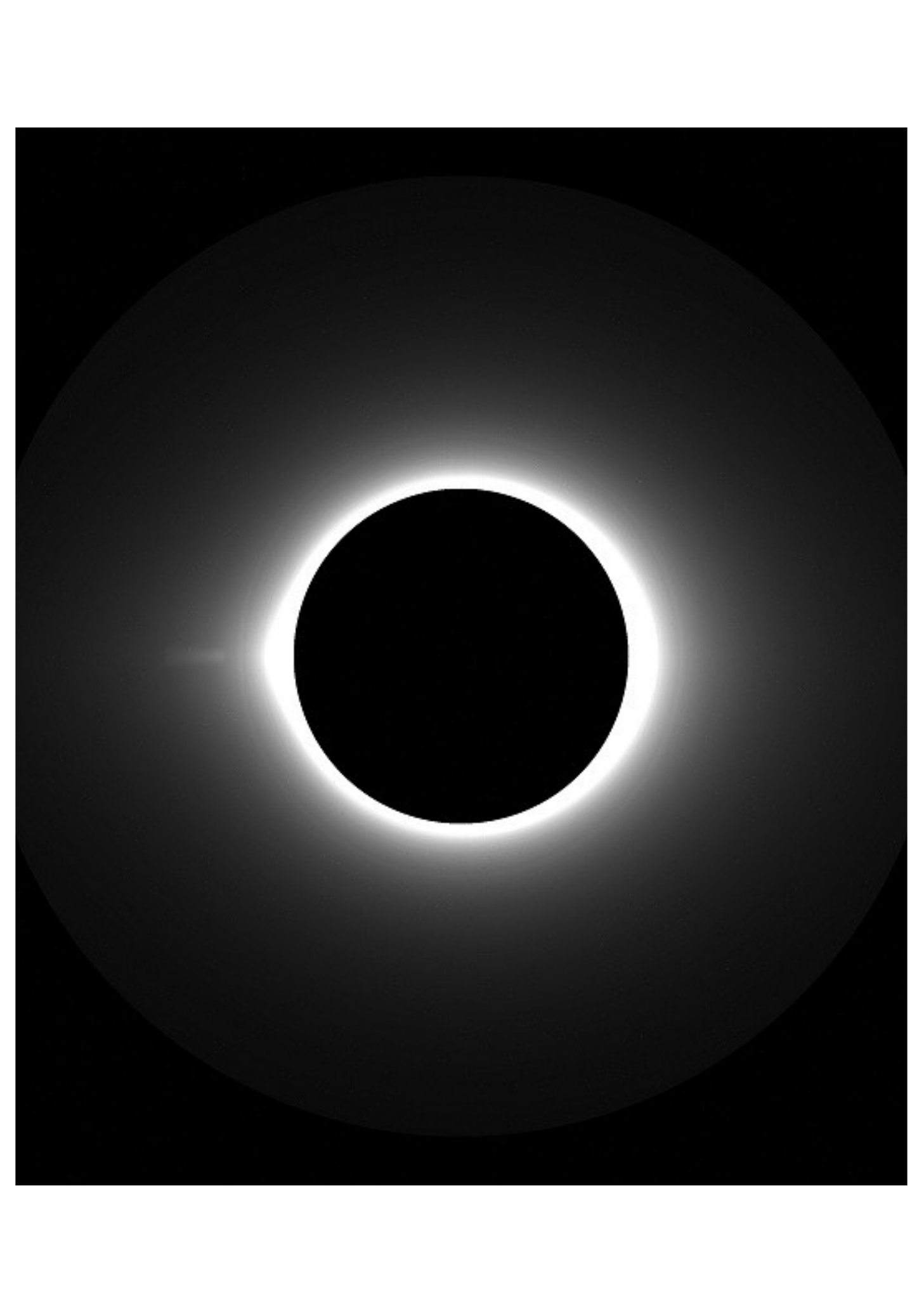}
              }
     \vspace{-0.1\textwidth}  
     \centerline{\Large     
      \hspace{0.43\textwidth}  \color{white}{(a)}
      \hspace{0.33\textwidth}  \color{white}{(b)}
         \hfill}
     \vspace{0.005\textwidth}     
          
 \centerline{\hspace*{0.05\textwidth}
               \includegraphics[width=0.39\textwidth,clip=]{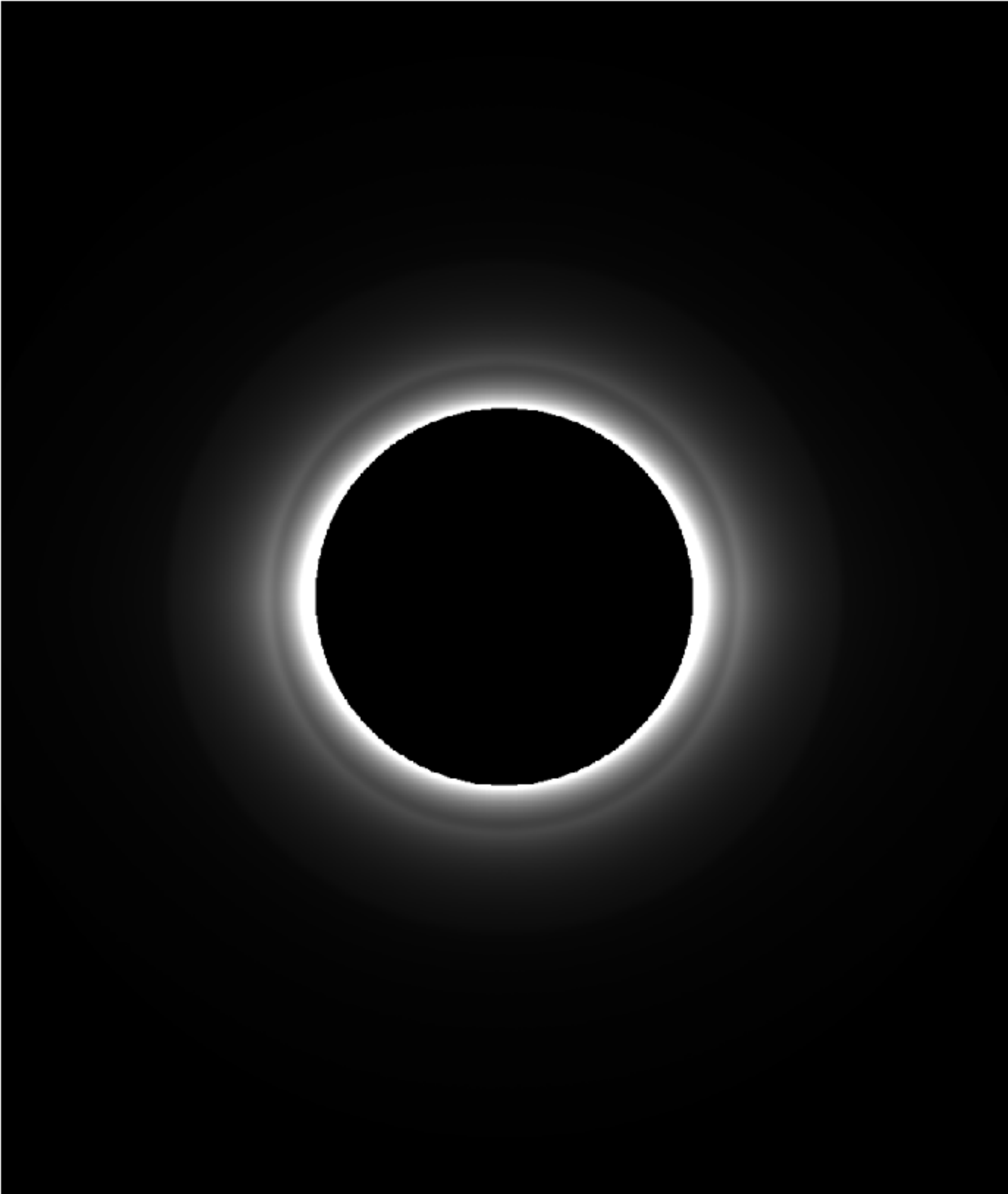}
               \hspace*{0.002\textwidth}
               \includegraphics[width=0.39\textwidth,clip=]{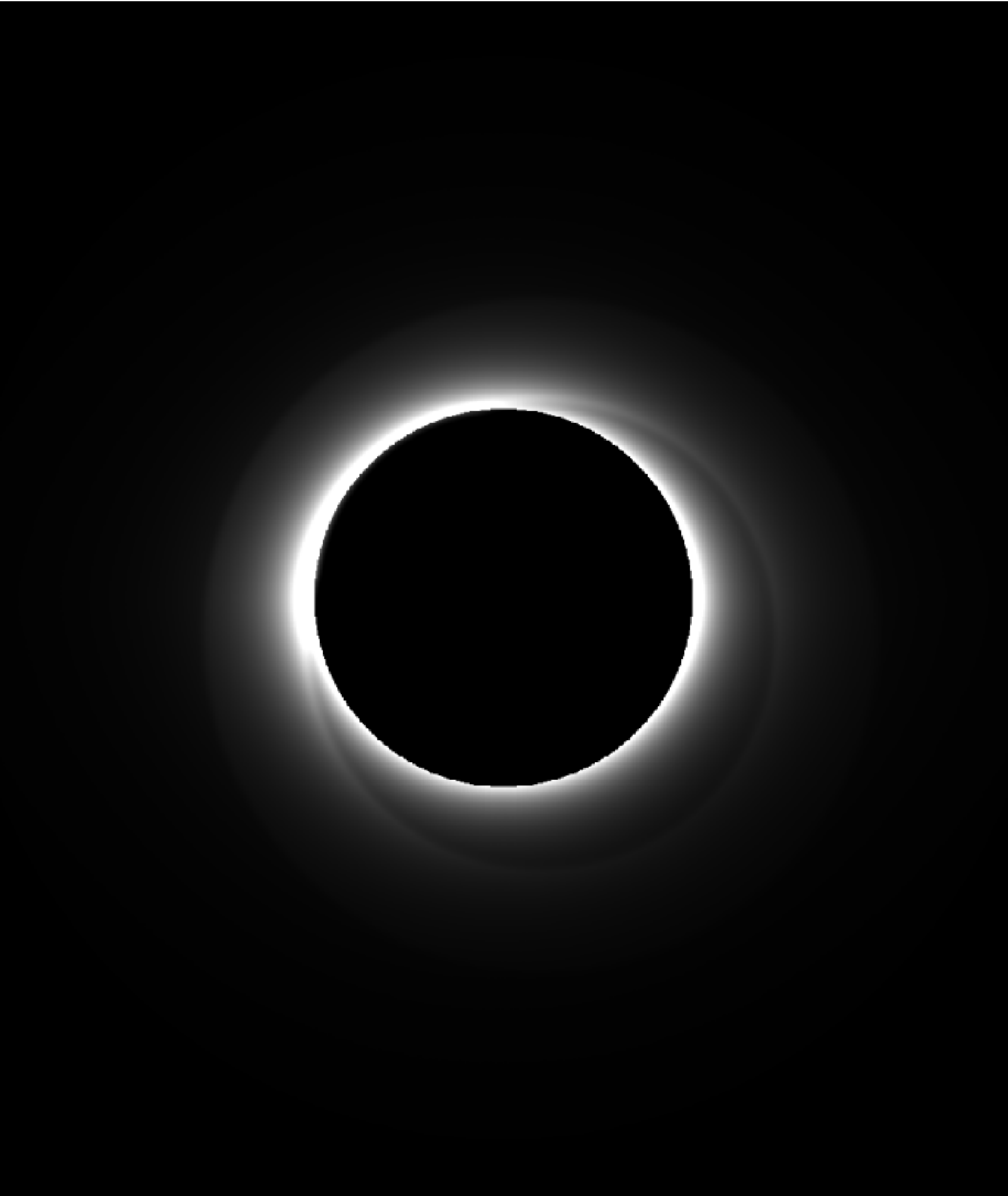}
              }
     \vspace{-0.05\textwidth}  
     \centerline{\Large     
      \hspace{0.43\textwidth}  \color{white}{(c)}
      \hspace{0.33\textwidth}  \color{white}{(d)}
         \hfill}
     \vspace{0.01\textwidth} 
          
\caption{{ Examples of simulated CMEs in VELC FOV, (a) Normal CME at west solar limb, (b) Narrow CME at east solar limb, (c) Symmetric halo CME, (d) Asymmetric halo CME, with 50\%, 50\%, 50\% and 60\% enhanced intensity with respect to background respectively. }
        }
   \label{CME eg}
\end{figure}

Figure \ref{CME eg}a shows the simulated normal type CME in VELC synthetic coronal image at 270$^{\circ}$ from solar north (measured anti-clockwise), i.e. west limb, with brightness enhancement of 50\% with respect to coronal brightness. Figure \ref{CME eg}b shows simulated narrow CME at east solar limb and having same brightness like that of normal CME of previous case.{ Figure \ref{CME eg}c is an example of symmetric halo CME while Figure \ref{CME eg}d is assymetric halo CME. Halo CMEs in the two cases have 50\% and 60\% intensity enhancement with respect to background respectively.} Scattered intensity has not been added in these examples to revel  the shape of CMEs, otherwise in presence of scatter light they will not be visible. 

\section{Detection Algorithm}
    \label{S-Algo}
\begin{figure}[!ht]
   \centerline{\includegraphics[width=1\textwidth,clip=]{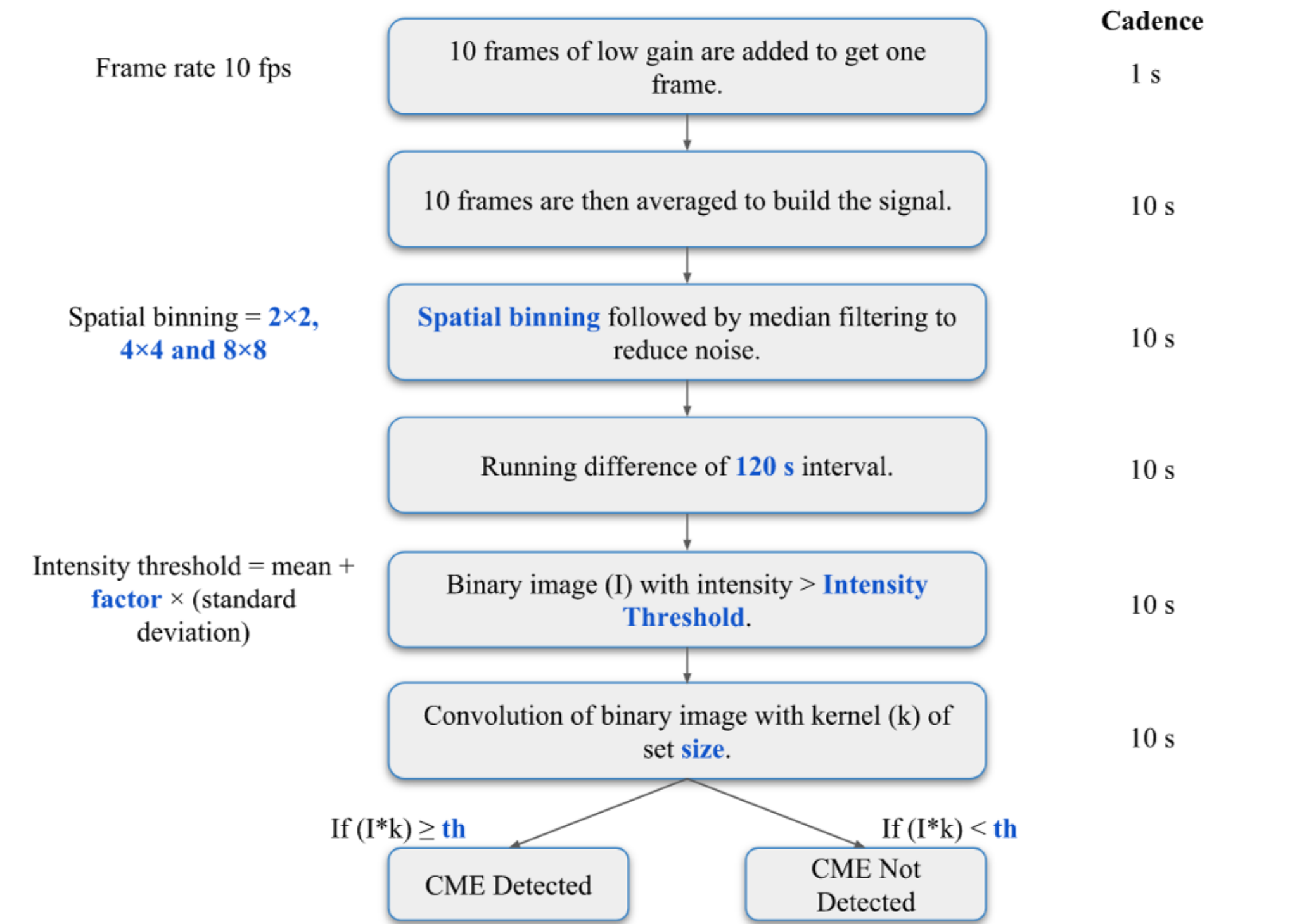}
              }
              \vspace{0.01\textwidth}
              \caption{Flow chart of automated CMEs detection algorithm developed for ADITYA-L1 VELC. The parameters shown in bold and blue letters are the free parameters.
                      }
   \label{flow chart}
\end{figure}

There are several automated CMEs detection algorithm (For example CACTus, SEEDS, ARTEMIS, CORIMP etc) that can detect CMEs in coronagraph images. However, these are on-ground algorithms which require huge amount of memory for the process. The automated CME detection algorithm of VELC is a simple algorithm that can be implemented in onboard electronics and is capable of detecting CMEs in coronagraph images by intensity thresholding with threshold value depending on the mean intensity of each image. It is followed by area thresholding with threshold value depending on the size of kernel used for the same. The flow chart in Figure \ref{flow chart} shows step by step application of the algorithm.  There are some free parameters in the algorithm which has been shown in bold and blue letters. The values of free parameters can be changed post launch for faithful detection of CMEs depending on the science requirements.

\begin{enumerate}[(i)]
\item Images are acquired at the rate of 10 frames per second (fps). The algorithm adds 10 images of 100 ms exposure to build the signal in first place. Thus, every second one image will be produced. 

\item  In order to increase the signal further, 10 such images of 1 s equivalent exposure are then averaged to obtain one image. The image so obtained will have improved signal to noise.

\item The noise in the images obtained in previous step is reduced by performing spatial binning in images followed by median filtering in temporal domain. The median filtering is to be performed over three images at a time. The binning size can be $2\times2$, $4\times4$ or $8\times8$ which is kept as free parameter for faithful detection of CMEs { i.e. the algorithm should be able to detect maximum number of images which actually contain CMEs and hence the amount of useful data lost is minimum. Also, the false detection should be less.} 

\item Running difference of the coronagraph images 120 s apart is taken to remove the F corona, instrument scattered light and less variable K corona. The interval 120 s is kept as free parameter in order to detect CMEs of variable speed. Range of this interval extends from 60 s (1 min) to 300 s (5 min) to detect the CMEs of all speeds. 

\item The images are then converted to binary images (I) by thresholding the intensity. The threshold value is defined as mean + factor$\times $(standard deviation)$ $. As the CMEs intensity and morphology varies widely, therefore, adaptive thresholding is applied such that the threshold value does not remain a fixed number for all images. Factor is a free parameter with values ranging from 1 to 3. 

\item The binary image so obtained is then convolved with a kernel (k) that has size as free parameter. The size can be $5\times5$, $10\times10$ or $20\times20$ depending on the size of CME to be detected based on science requirements. 

\item In the convolved image, the pixels are checked with a convolution threshold (th) that determines the presence of CME in an image. Too low threshold can lead to detection of noise, whereas too high value will detect only big and bright CMEs. Therefore, it is kept as free parameter which can be tuned post launch. If the maximum value in the convolved image is more than the convolution threshold, then CME is said to be detected in the image, else it stops detection. A range of 0.4-1 can been considered for optimal detection.

\item If the binary image satisfies both the threshold, then steps 1-7 are applied to the next image. The algorithm checks 5 consecutive frames for CMEs to avoid false detection before storing in the memory. The algorithm continues to run and detect CMEs in successive images until no CME is detected. When the detection stops, the images are added to form one image if there is no CME for 3 hours or till the next detection, whichever is earlier.
\end{enumerate}

This algorithm retain images containing CMEs while discarding the rest. The performance of this algorithm has been discussed in the following sections using various data sets.

\section{Application to Existing Coronagraph Images}
\label{S-Results existing} 

The efficiency of the application of algorithm on different data sets is judged on the basis of the following quantities which has been calculated at each threshold.

\begin{itemize}
    \item Relative CMEs detection (RCD) is the percentage for CMEs images that has been detected by the algorithm,
    \begin{equation}
        \mathrm{{RCD}} = \frac{\mathrm{{Number\; of\; CME\; images\; detected}}}{\mathrm{{Total\; number\; of\; CME\; images}}}\times 100.
    \end{equation}
    \item Absolute CMEs detection (ACD) is calculated as percentage of total number of images in a day which actually contain CMEs and has been detected. This quantity is important for estimation of telemetry reduction,
    \begin{equation}
       \mathrm{{ACD}} = \frac{\mathrm{{Number\; of\; detected\; images\; actually\; containing\; CMEs}}}{\mathrm{{Total\; number\; of\; images}}}\times 100.
    \end{equation}
    \item Extra detected (ED) which can also be called as false detection is calculated as percentage of total number of images detected to contain CMEs which do not actually contain CMEs,
    \begin{equation}
       \mathrm{{ED}} = \frac{\mathrm{{Number\; of\; images\; falsely\; detected}}}{\mathrm{{Number\; of\; positively\; detected\; images}}}\times 100.
    \end{equation}
    \item Reduced Telemetry (RT) percentage is calculated as percentage of total images which are detected for CMEs. These images contain all images with and without CMEs detected by the algorithm,
    \begin{equation}
       \mathrm{{RT}} = \frac{\mathrm{{Positively\; Detected\; images}}}{\mathrm{{Total\; number\; of\; images}}}\times 100.
    \end{equation}
\end{itemize}

\subsection{Application to K-Cor Data}
    \label{S-kcor}
    
The K-Cor telescope (\citealp{deWijn}) is one of the many instruments in Mauna Loa Solar Observatory (MLSO). It is a coronagraph with FOV same as that of VELC starting from 1.05 R$_{\odot}$ to 3 R$_{\odot}$.
K-Cor takes images of corona with dimension of $1024\times1024$ and cadence of 15 s. The level 1 images were downloaded from the website \url{http://mlso.hao.ucar.edu/calendars/mlso_data_KCOR_2015.php}. Since it is ground based instrument it suffers from atmospheric effects. We have tested our algorithm on some of the images of K-Cor and got positive results. Though it could not be applied on any data-set, so we selected some good events in which CMEs can be visually seen and some events which are affected by atmosphere.
The images were binned by $2\times2$ pixels and median filtering was done in temporal domain taking three images at a time to reduce noise. Running difference of images 5 minutes apart were then taken and binary image of the difference image was made taking factor as 1, 1.5 and 2. The resulting image was convolved with the kernel of size $10\times10$ pixels. The presence of CME was detected by comparing the maximum of the convolved image with convolution threshold value of 0.6.  

\begin{figure}[!ht]   
   \centerline{\hspace*{0.015\textwidth}
               \includegraphics[width=0.6\textwidth,clip=]{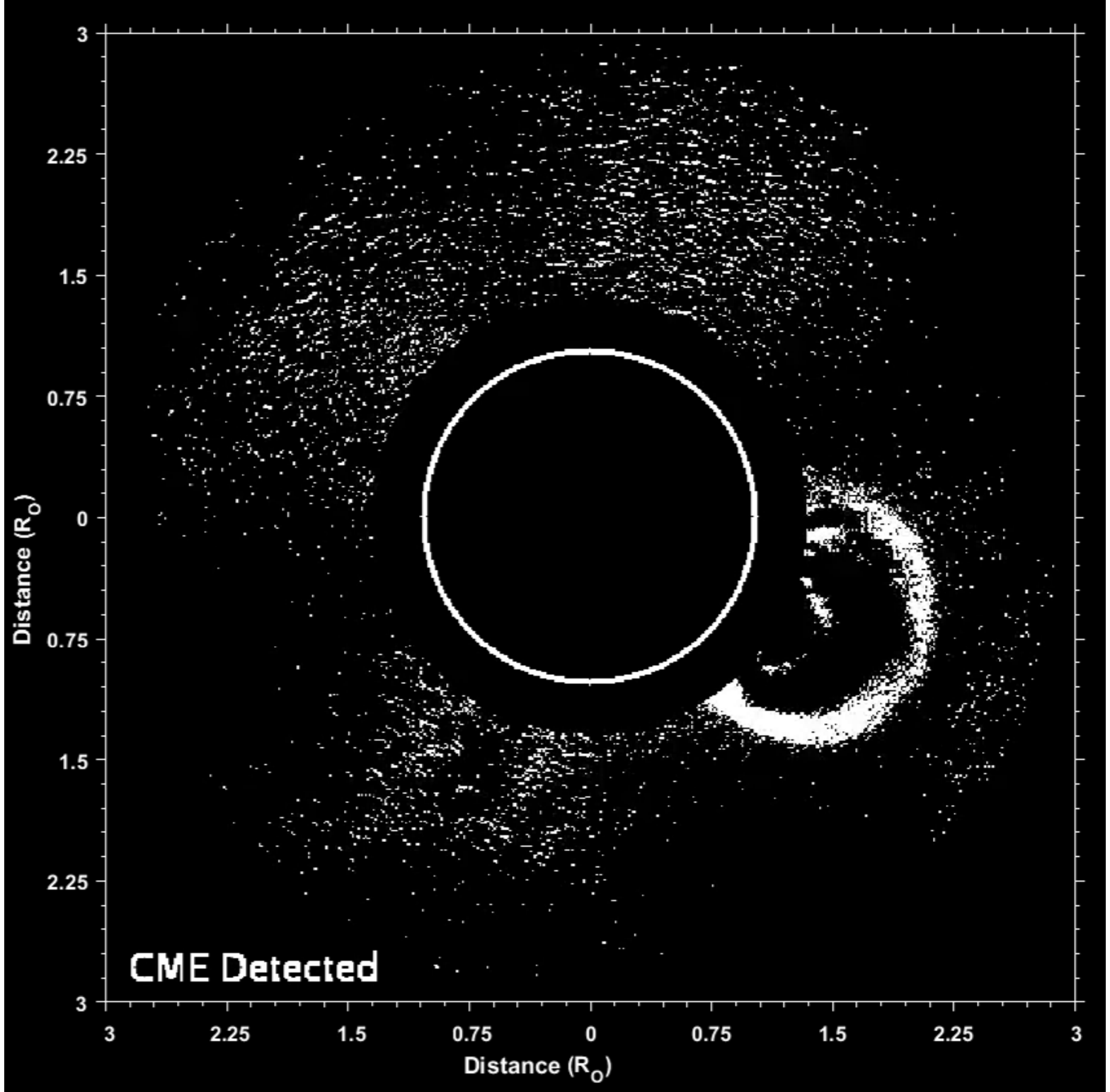}
               \hspace*{-0.055\textwidth}
               \includegraphics[width=0.6\textwidth,clip=]{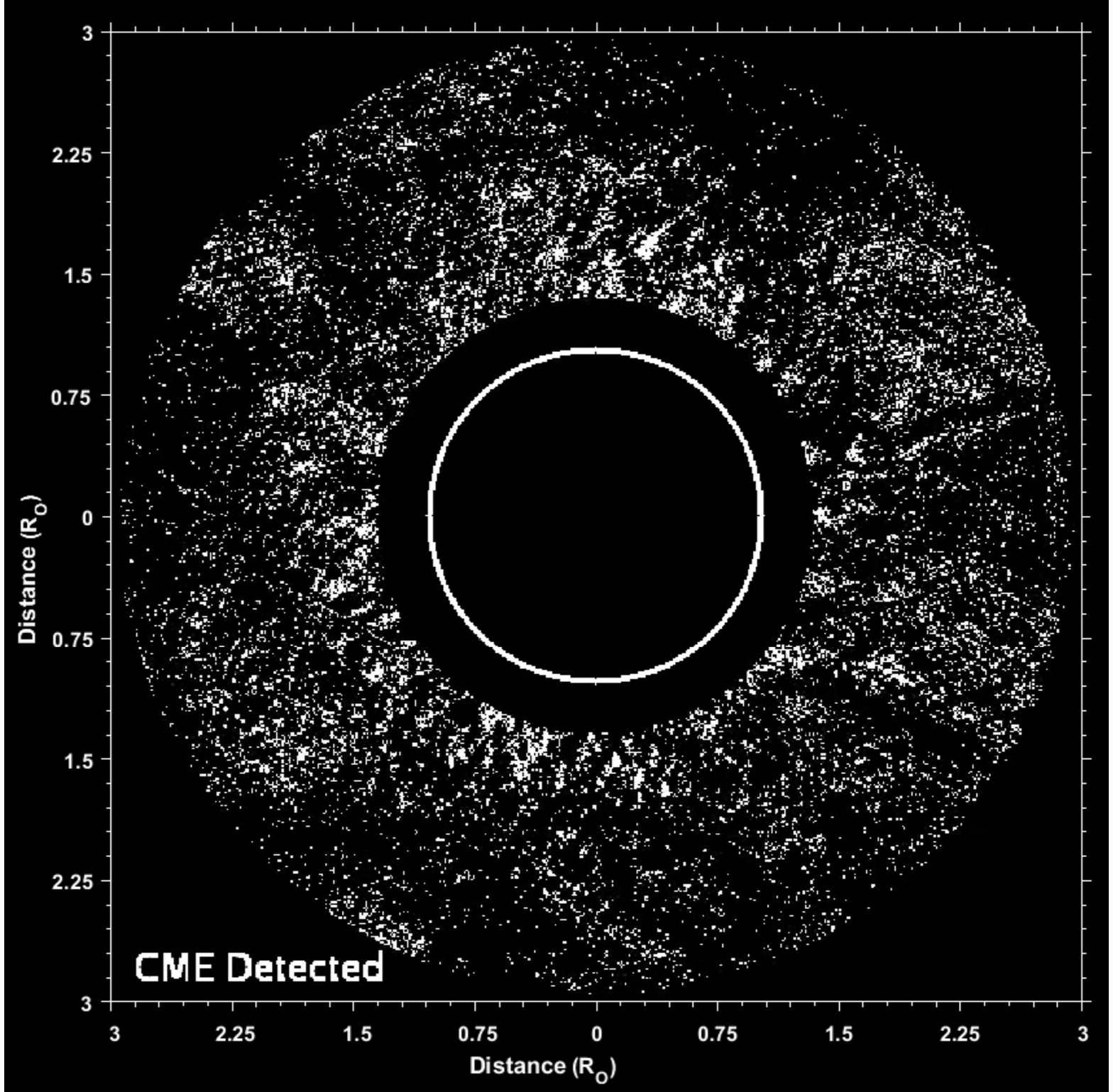}
              }
     \vspace{0.015\textwidth}   
     
 \caption{\textit{Left:} CME detection in K-cor image of 2016-01-01. The bright leading edge and less atmospheric effects clearly shows the detected CME. \textit{Right:} A false detection of CME is shown in K-Cor image of 2014-05-28 on account of huge atmospheric contributions.
        }
   \label{kcor Application}
\end{figure}

Figure \ref{kcor Application}a shows an example of good CME event captured by K-cor on 2016-01-01 and was successfully detected by the algorithm. However, for most of the cases, the algorithm fails to detect CME alone in K-cor images as the atmospheric effects contributes to the scattered intensities which resulted in false detection by the algorithm. Such an example is show in Figure \ref{kcor Application}b captured by K-cor on 2014-05-28. Table \ref{kcor table} summarizes the results after the application of the algorithm to K-Cor data. There are 501 images on 09-02-2016 which actually contains CMEs and has been detected. The value of RCD is 80.4\% which is the ratio of images detected containing CMEs (501) and number of images containing CMEs (623). ACD is the ratio of CMEs images detected (501) to the total number of images for this day (1709). Out of the total positively detected images (1041), the number of falsely detected images is 540. This ratio gives the extra detection which is 51.8\% in this case. RT is calculated taking the ratio of positive detected images (1041) and total number of images for that day (1709), which comes out as 60.9\%. The number of detected CMEs images and extra detected has not been explicitly mentioned in the tables. All the quantities for rest of the cases has been calculated in similar way.

\begin{sidewaystable}
\caption{Application of the algorithm to K-Cor data taking kernel size $10\times10$ and convolution threshold of 0.6.}
\label{kcor table}
\begin{tabular}{cccccccccc}
\hline
S. No. & \begin{tabular}[c]{@{}c@{}}Date and\\[0.5ex] total number \\[0.5ex] of images\end{tabular} & \begin{tabular}[c]{@{}c@{}}Number \\[0.5ex] of CME\\[0.5ex] images\end{tabular} & \begin{tabular}[c]{@{}c@{}}Factor\\[0.5ex] (f)\end{tabular} & \begin{tabular}[c]{@{}c@{}}Positive\\[0.5ex] Detected\\[0.5ex] Images\end{tabular} & \begin{tabular}[c]{@{}c@{}}Relative\\[0.5ex] CMEs\\[0.5ex] Detected\\[0.5ex] RCD (\%)\end{tabular} & \begin{tabular}[c]{@{}c@{}}Absolute\\[0.5ex] CMEs\\[0.5ex] detected\\[0.5ex] ACD (\%)\end{tabular} & \begin{tabular}[c]{@{}c@{}}Extra\\[0.5ex] detected\\[0.5ex] ED (\%)\end{tabular} & \begin{tabular}[c]{@{}c@{}}Reduced\\[0.5ex] Telemetry\\[0.5ex] RT (\%)\end{tabular} & Remarks \\ \hline
\multirow{3}{*}{1} & \multirow{3}{*}{\begin{tabular}[c]{@{}c@{}}2016-02-09\\[1ex] 1709\end{tabular}} & \multirow{3}{*}{623} & 1 & 1041 & 80.4 & 29.3 & 51.8 & 60.9 & \multirow{3}{*}{Bright CME} \\ 
 &  &  & 1.5 & 592 & 62.4 & 22.7 & 34.2 & 34.6 &  \\ 
 &  &  & 2 & 335 & 41.4 & 15.1 & 22.9 & 19.6 &  \\ \hline
\multirow{3}{*}{2} & \multirow{3}{*}{\begin{tabular}[c]{@{}c@{}}2016-01-01\\[1ex] 284\end{tabular}} & \multirow{3}{*}{155} & 1 & 164 & 58.06 & 31.6 & 45.1 & 57.7 & \multirow{3}{*}{\begin{tabular}[c]{@{}c@{}}Very Bright\\[0.5ex] CME\end{tabular}} \\ 
 &  &  & 1.5 & 107 & 44.5 & 24.2 & 35.5 & 37.6 &  \\ 
 &  &  & 2 & 55 & 24.5 & 13.3 & 30.9 & 19.3 &  \\ \hline
\multirow{3}{*}{3} & \multirow{3}{*}{\begin{tabular}[c]{@{}c@{}}2015-05-05\\[1ex] 1293\end{tabular}} & \multirow{3}{*}{405} & 1 & 601 & 42.4 & 13.3 & 71.3 & 46.5 & \multirow{3}{*}{\begin{tabular}[c]{@{}c@{}}Bright CME.\\[0.5ex] Leading edge\\[0.5ex] missed\end{tabular}} \\ 
 &  &  & 1.5 & 246 & 26.6 & 8.3 & 56.1 & 19 &  \\ 
 &  &  & 2 & 72 & 15 & 4.7 & 15.2 & 5.5 &  \\ \hline
\multirow{3}{*}{4} & \multirow{3}{*}{\begin{tabular}[c]{@{}c@{}}2014-06-26\\[1ex] 1762\end{tabular}} & \multirow{3}{*}{134} & 1 & 32 & 20.1 & 1.5 & 15.6 & 1.8 & \multirow{3}{*}{Faint CME} \\ 
 &  &  & 1.5 & 7 & 5.2 & 0.39 & 0 & 0.39 &  \\ 
 &  &  & 2 & 0 & 0 & 0 & 0 & 0 &  \\ \hline
 \multirow{3}{*}{5} & \multirow{3}{*}{\begin{tabular}[c]{@{}c@{}}2014-05-28\\[1ex] 1582\end{tabular}} & \multirow{3}{*}{353} & 1 & 555 & 36.2 & 8.1 & 76.9 & 35.1 & \multirow{3}{*}{\begin{tabular}[c]{@{}c@{}}Medium Bright\\[0.5ex] One CME cannot be \\[0.5ex] distinctly seen.\end{tabular}} \\ 
 &  &  & 1.5 & 231 & 26.8 & 3.9 & 73.1 & 14.6 &  \\
 &  &  & 2 & 85 & 40.0 & 2.15 & 60.0 & 5.3 &  \\ \hline
 \multirow{3}{*}{6} & \multirow{3}{*}{\begin{tabular}[c]{@{}c@{}}2014-05-07\\[1ex] 1149\end{tabular}} & \multirow{3}{*}{78} & 1 & 949 & 80.7 & 5.4 & 93.3 & 82.5 & \multirow{3}{*}{\begin{tabular}[c]{@{}c@{}}Very faint.\\[0.5ex] CME cannot be \\[0.5ex] distinctly seen.\end{tabular}} \\ 
 &  &  & 1.5 & 567 & 51.2 & 3.4 & 92.9 & 49.3 &  \\
 &  &  & 2 & 257 & 7.6 & 0.52 & 97.6 & 22.3 &  \\ \hline
\multirow{3}{*}{7} & \multirow{3}{*}{\begin{tabular}[c]{@{}c@{}}2014-04-03\\[1ex] 1615\end{tabular}} & \multirow{3}{*}{422} & 1 & 776 & 97.4 & 25.4 & 47.03 & 48.04 & \multirow{3}{*}{\begin{tabular}[c]{@{}c@{}}Very faint.\\[0.5ex] CME cannot be \\[0.5ex] distinctly seen\end{tabular}} \\ 
 &  &  & 1.5 & 518 & 90.2 & 23.6 & 26.2 & 32.07 &  \\ 
 &  &  & 2 & 290 & 57.5 & 15.04 & 16.2 & 17.9 &  \\ \hline
\end{tabular}
\end{sidewaystable}

\begin{itemize}
    \item CMEs which can be visually seen in difference images are detected with considerable efficiency. These include the events of 2016-02-09, 2016-01-01 and 2015-05-05. 
    \item Faint CMEs are hard to detect on account of atmospheric contribution. The best being only 1.8\% of images which were detected for CMEs in this case.
    \item There are two cases of very faint CMEs in which CMEs cannot be identified manually but the algorithm showed presence of CMEs in images of dates 2014-05-07 and 2014-04-03. After careful observation and comparison with LASCO CME database, the position of CMEs were identified. Though very faint propagating structures are visible, the large contribution of atmospheric scattered intensity was responsible for false detection. Figure \ref{kcor Application}b shows such an example of false detection.
    \item Since, K-Cor is ground based coronagraph, there is high percentage of false detection. Almost half of the images detected do not contain CMEs.
    \item The average telemetry for the cases with considerable CMEs detection are $\approx$55\%, $\approx$30\% and $\approx$15\% where as for very faint cases it is $\approx$65\%, $\approx$41\% and $\approx$20\% for factors of 1, 1.5 and 2 respectively.
\end{itemize}

\subsection{Application to STEREO COR-1A Data}
     \label{S-cor1}
     
Among the existing space-based coronagraphs, FOV of STEREO COR-1A is nearest to that of VELC. So, this data set has been used to test the detection algorithm. STEREO COR-1A polarized brightness images are acquired in triplets with polarization angles of 0$^{\circ}$, 120$^{\circ}$ and 240$^{\circ}$ with respect to the solar north. Each of these images is taken at an interval of 9 or 12 s, and every triplet has difference in time of 5 minutes. Therefore, everyday 864 polarized images are acquired, which corresponds to 288 polarization triplets each day. These images are taken with an exposure time of $\approx$1.7 s and are of dimension $512\times512$ pixels. The STEREO COR-1A observation log specifies high jitter data while the other data sets can be treated as low jitter data. For this work we selected the images from dates { 2011-09-22,} 2011-05-14, 2011-05-15, 2012-04-17, 2012-04-19, 2013-10-26, 2013-10-27, 2013-10-28 which contain high jitters and different types of CMEs. Images from dates 2010-11-10, 2010-11-11, 2010-11-12, 2011-08-02, 2011-08-03, 2011-08-04, 2012-06-13, 2012-06-14, 2012-06-15 were selected for low jitters and different types of CMEs.
The standard routines in SECCHI package provided in SolarSoftware has been used to calibrate each image and then each triplet of polarized images has been combined to derive the total brightness (tB) image.

The algorithm is implemented on total brightness images. Median filtering is applied at each pixel of every three consecutive images to reduce noise. Then running difference images are obtained from these consecutive median images. The factor for intensity threshold is taken as 1.5. The binary image generated after intensity threshold is then convolved with kernel of size $8\times8$. The convolution threshold is varied as 0.4, 0.5 and 0.6 and the performance of the algorithm is analysed.

\begin{figure}[!ht]   
  \centerline{\hspace*{0.015\textwidth}
              \includegraphics[width=0.6\textwidth,clip=]{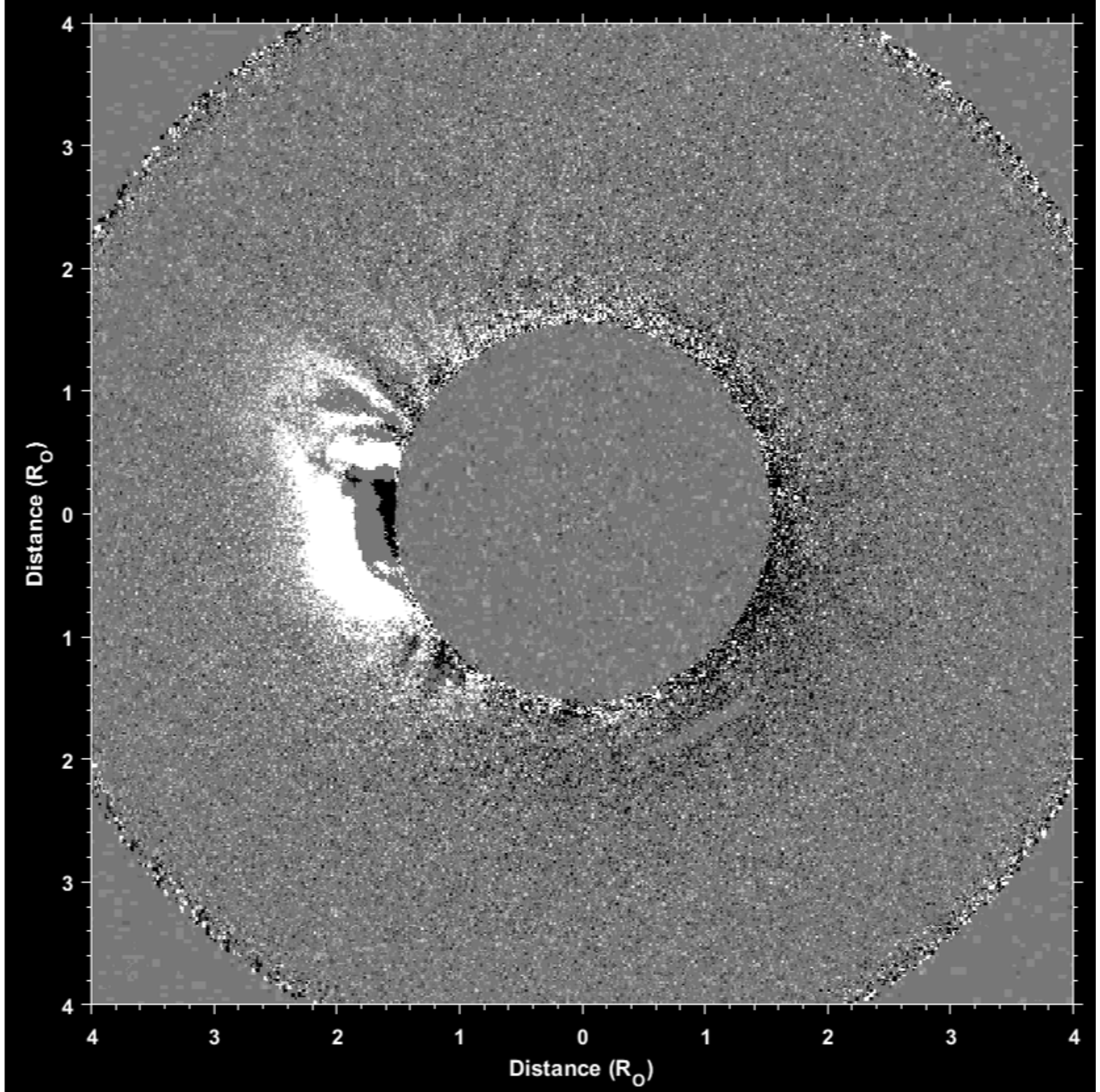}
              \hspace*{-0.15\textwidth}
              \includegraphics[width=0.6\textwidth,clip=]{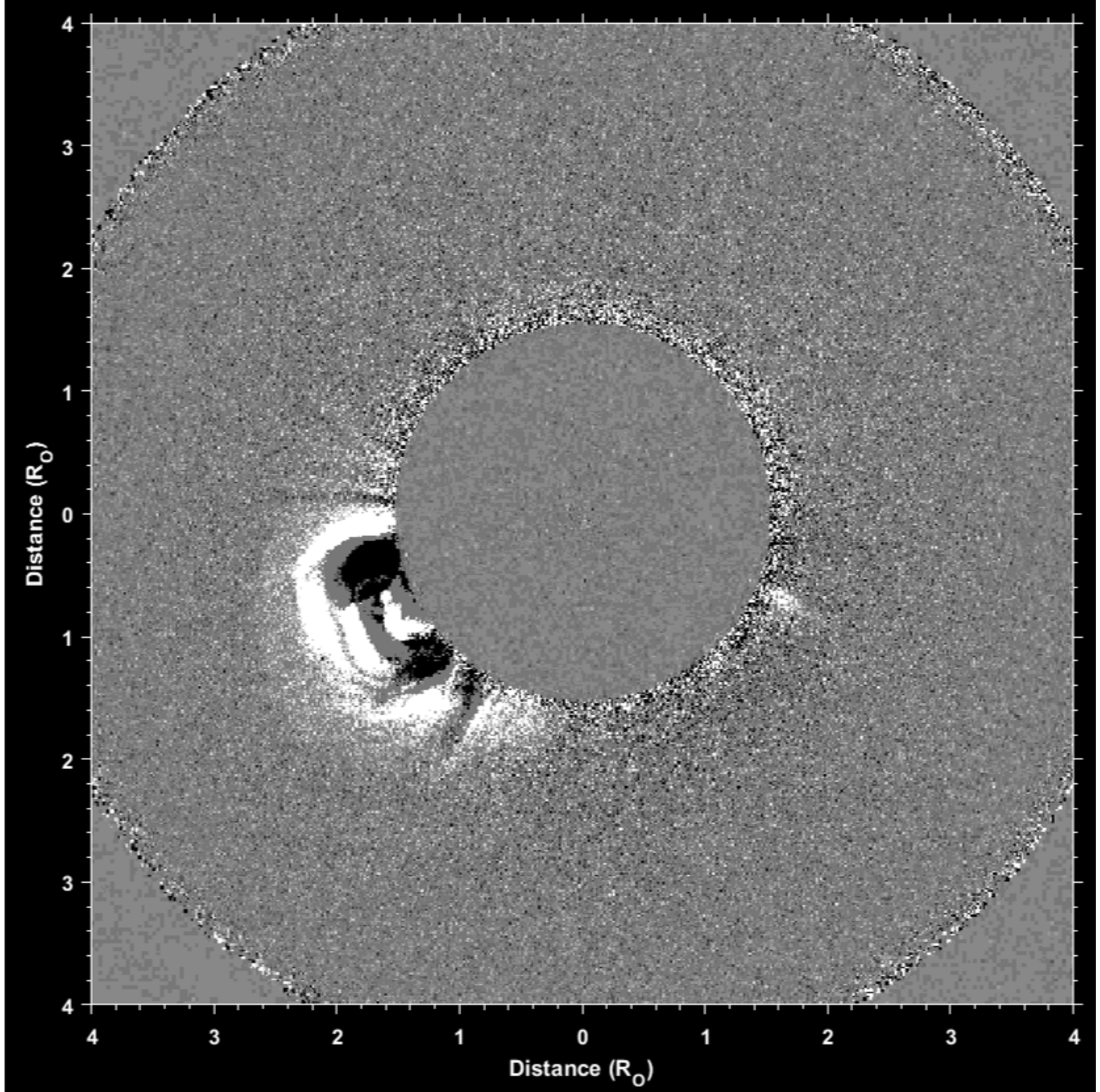}
              }
     \vspace{-0.08\textwidth}  
    
     \centerline{\Large     
      \hspace{0.405 \textwidth}  \color{white}{(a)}
      \hspace{0.375\textwidth}  \color{white}{(b)}
         \hfill}
     \vspace{0.045\textwidth}     
          
  \centerline{\hspace*{0.015\textwidth}
              \includegraphics[width=0.6\textwidth,clip=]{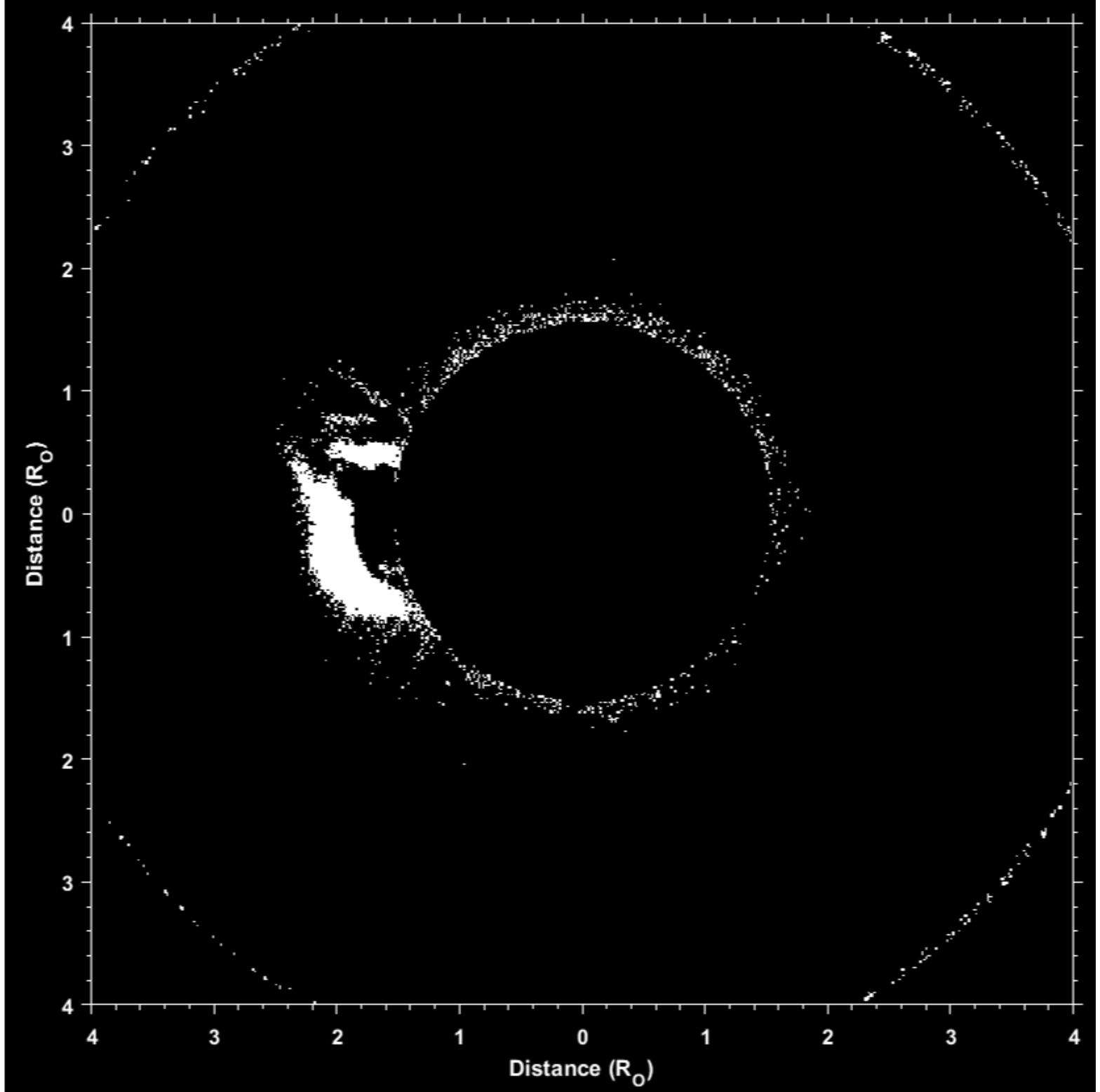}
              \hspace*{-0.15\textwidth}
              \includegraphics[width=0.6\textwidth,clip=]{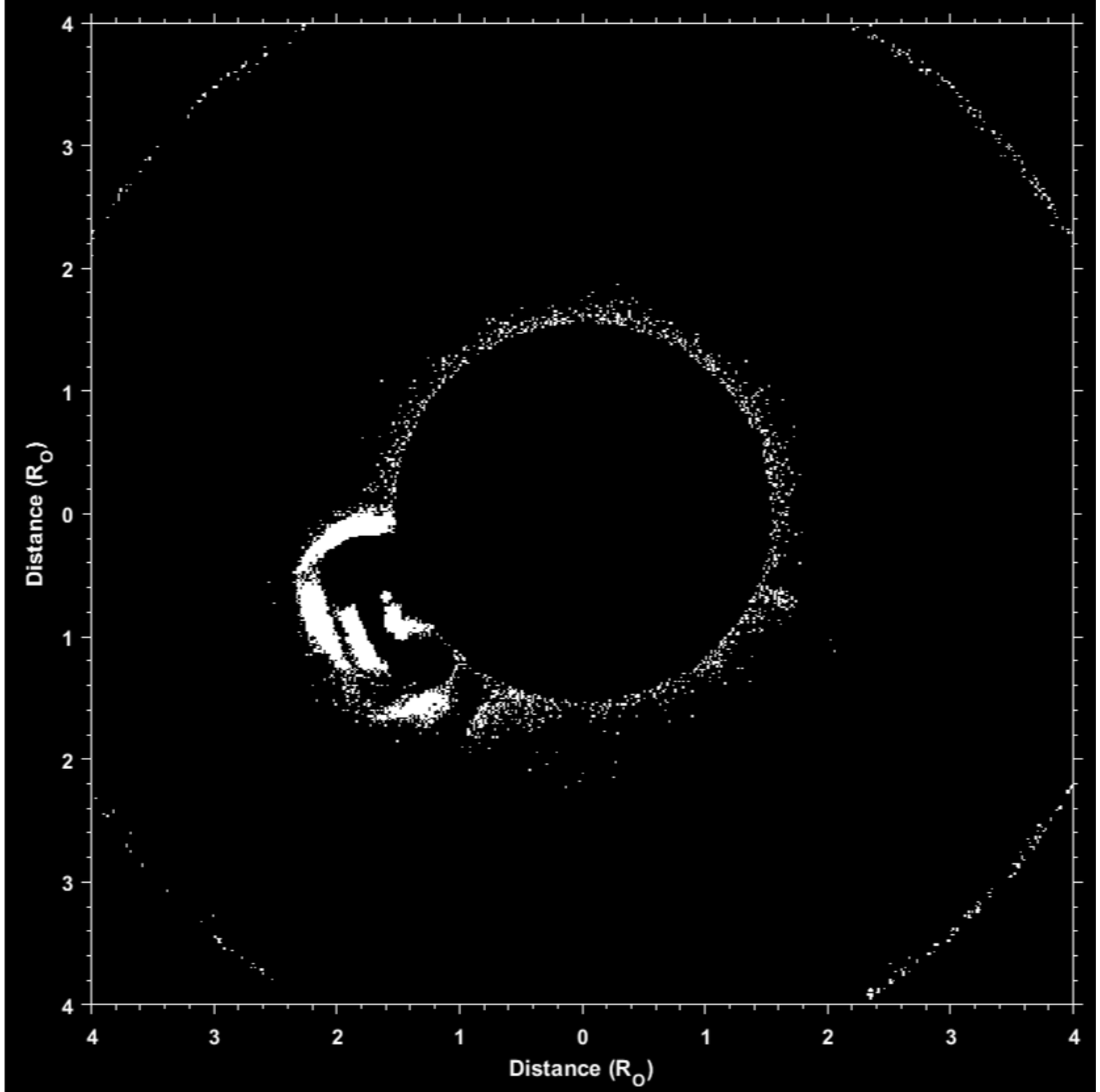}
              }
     \vspace{-0.08\textwidth}    
     \centerline{\Large     
      \hspace{0.405 \textwidth}  \color{white}{(c)}
      \hspace{0.375\textwidth}  \color{white}{(d)}
         \hfill}
     \vspace{0.05\textwidth}    
          
\caption{COR-1A images after application of the algorithm. The upper panel shows running difference images while the lower panel shows binary images in which CMEs are detected. The left column in both panels correspond to high jitter case for date 2013-10-28 while the right column correspond to low jitter case for date 2012-06-14.}
  \label{COR1 imgs}
\end{figure}

Figure \ref{COR1 imgs} shows result of application of the algorithm on COR-1A images for dates 2013-10-28 and 2012-06-14. These two examples correspond to high jitter and low jitter cases respectively. It can be seen that most part of CMEs in both cases has been detected after the binary image is produced of running difference images. The results of all test cases after the application of algorithm with different values of thresholds for high and low jitter cases are summarized in Table \ref{high jitter table} and \ref{low jitter table} respectively.

\subsubsection{High jitter}

\begin{sidewaystable}[]
\centering
\caption{Application of CMEs Detection algorithm to COR-1A images for high jitter taking kernel of size $8\times8$ and factor for intensity threshold as 1.5}
\label{high jitter table}
\begin{tabular}{cccccccccc}
\hline
S. No. & \begin{tabular}[c]{@{}c@{}}Date\\[0.5ex] and total\\[0.5ex] number of\\[0.5ex] images\end{tabular} & \begin{tabular}[c]{@{}c@{}}Number \\[0.5ex] of CMEs\\[0.5ex]images\end{tabular} & \begin{tabular}[c]{@{}c@{}}Convolution\\[0.5ex] threshold\end{tabular} & \begin{tabular}[c]{@{}c@{}}Positive\\[0.5ex] detected \\[0.5ex] images\end{tabular} & \begin{tabular}[c]{@{}c@{}}Relative\\[0.5ex] CMEs \\[0.5ex] detected\\[0.5ex] RCD (\%)\end{tabular} & \begin{tabular}[c]{@{}c@{}}Absolute\\[0.5ex] CMEs \\[0.5ex] detected\\[0.5ex] ACD (\%)\end{tabular} & \begin{tabular}[c]{@{}c@{}}Extra\\[0.5ex] Detected\\[0.5ex] ED (\%)\end{tabular} & \begin{tabular}[c]{@{}c@{}}Reduced\\[0.5ex] telemetry\\[0.5ex] RT (\%)\end{tabular} & Remarks \\ \hline
\multirow{3}{*}{1} & \multirow{3}{*}{\begin{tabular}[c]{@{}c@{}}2011-05-14\\[1ex] 284\end{tabular}} & \multirow{3}{*}{8} & 0.4 & 211 & 50 & 2.11 & 97.15 & 74.29 & \multirow{3}{*}{\begin{tabular}[c]{@{}c@{}}Very faint\\[0.5ex] CME\end{tabular}} \\
 &  &  & 0.5 & 45 & 25 & 0.7 & 95.55 & 15.84 &  \\
 &  &  & 0.6 & 25 & 12.5 & 0.35 & 96 & 8.80 &  \\ \hline
\multirow{3}{*}{2} & \multirow{3}{*}{\begin{tabular}[c]{@{}c@{}}2011-05-15\\[1ex] 284\end{tabular}} & \multirow{3}{*}{59} & 0.4 & 263 & 98.3 & 20.42 & 77.94 & 92.60 & \multirow{3}{*}{Faint CME} \\
 &  &  & 0.5 & 134 & 74.57 & 15.49 & 67.16 & 47.18 &  \\
 &  &  & 0.6 & 82 & 54.23 & 11.26 & 60.97 & 28.87 &  \\ \hline
\multirow{3}{*}{3} & \multirow{3}{*}{\begin{tabular}[c]{@{}c@{}}2012-04-17\\[1ex] 257\end{tabular}} & \multirow{3}{*}{70} & 0.4 & 225 & 92.85 & 25.29 & 71.11 & 87.54 & \multirow{3}{*}{\begin{tabular}[c]{@{}c@{}}Increased\\[0.5ex] jitter\end{tabular}} \\
 &  &  & 0.5 & 96 & 57.14 & 15.56 & 58.33 & 37.35 &  \\
 &  &  & 0.6 & 46 & 37.14 & 10.11 & 43.47 & 17.89 &  \\ \hline
\multirow{3}{*}{4} & \multirow{3}{*}{\begin{tabular}[c]{@{}c@{}}2012-04-19\\[1ex] 278\end{tabular}} & \multirow{3}{*}{68} & 0.4 & 261 & 100 & 24.46 & 73.94 & 93.88 & \multirow{3}{*}{Bright CMEs} \\
 &  &  & 0.5 & 129 & 85.29 & 20.86 & 55.03 & 46.40 &  \\
 &  &  & 0.6 & 74 & 73.52 & 17.98 & 32.43 & 26.61 &  \\ \hline
\multirow{3}{*}{5} & \multirow{3}{*}{\begin{tabular}[c]{@{}c@{}}2013-10-26\\[1ex] 2013-10-27\\[1ex] 2013-10-28\\[1ex] \\ 655\end{tabular}} & \multirow{3}{*}{170} & 0.4 & 651 & 90.58 & 25.80 & 74.03 & 99.38 & \multirow{3}{*}{\begin{tabular}[c]{@{}c@{}}Very faint\\[0.5ex] to bright CMEs\\[0.5ex] during this time\end{tabular}} \\
 &  &  & 0.5 & 616 & 88.23 & 25.19 & 73.21 & 94.04 &  \\
 &  &  & 0.6 & 533 & 81.17 & 23.05 & 71.48 & 81.37 & \\ \hline
 \multirow{3}{*}{{ 6}} & \multirow{3}{*}{\begin{tabular}[c]{@{}c@{}}{ 2011-09-22}\\[1ex] { 286}\end{tabular}} & \multirow{3}{*}{{ 26}} & { 0.4} & { 88} & { 92.31} & { 8.39} & { 72.72} & { 30.77} & \multirow{3}{*}{\begin{tabular}[c]{@{}c@{}}{ Faint CME}\\[0.5ex] { and a bright}\\[0.5ex] { halo CME}\end{tabular}} \\
 &  &  & { 0.5} & { 45} & { 84.61} & { 7.69} & { 51.11} & { 15.73} &  \\
 &  &  & { 0.6} & { 34} & { 73.07} & { 6.64} & { 44.11} & { 11.89} &  \\ \hline
\end{tabular}
\end{sidewaystable}

We tested the algorithm on several very faint, faint and bright CMEs for high jitter data. The results of table \ref{high jitter table} are 
\begin{itemize}
\item All CMEs identified manually are also identified by the automated algorithm.
\item For very faint, detection efficiency is found to be $\approx$50$\%$, 25$\%$ and 12$\%$ for convolution thresholds of 0.4, 0.5 and 0.6 respectively. 
\item For faint event, detection efficiency is $\approx$ 90$\%$, 50$\%$ and 30$\%$ for convolution thresholds of 0.4, 0.5 and 0.6 respectively except for the events occurred on 2013-10-26,27 and 28. It is because the STEREO spacecraft suffered high jitter on 2013-10-26,27 and 28. Due to which there are spurious brightness in FOV which is captured as CMEs by the algorithm.
\item For bright events, detection efficiency is $\approx$100$\%$, 81$\%$ and 81$\%$ for convolution thresholds of 0.4, 0.5 and 0.6 respectively except for the events occurred on 2013-10-26,27 and 28.
\item The average data volume for convolution threshold of 0.4, 0.5 and 0.6 is $\approx$85$\%$, 35$\%$ and 17$\%$ respectively. The data volume on 2013-10-26,27 and 28 is 99$\%$, 94$\%$ and 81$\%$  respectively. It suggest that almost all the images were detected to contain CMEs. It happened because of spurious brightness in FOV due to high jitter
\item Again we find that except for very faint events, the threshold of 0.5 is good enough to capture the CMEs and at the same time is able to reduce data volume to just 20$\%$. 
\item { We find that $\approx$92\%, 84\%, 73\% of halo CME images are also efficiently detected by the algorithm with reduced telemetry of $\approx$30\% and better for convolution thresholds of 0.4, 0.5 and 0.6 respectively.}
\item We also find that the algorithm is not affected much due to the presence of low jitter and high jitter. Whereas, for very high jitter the algorithm is not robust. It is because it depends on the brightness in the images.
Further reduction by a factor of 2 can be achieved by rice compression.
\end{itemize}

\subsubsection{Low jitter}

\begin{sidewaystable}[]
\caption{Application of CMEs Detection algorithm to COR-1A images for low jitter taking kernel of size $8\times8$ and factor for intensity threshold as 1.5}
\label{low jitter table}
\begin{tabular}{cccccccccc}
\hline
S.No. & \begin{tabular}[c]{@{}c@{}}Date\\[0.5ex] and total\\[0.5ex] number of\\[0.5ex] images\end{tabular} & \begin{tabular}[c]{@{}c@{}}Number \\[0.5ex] of CMEs \\[0.5ex]images\end{tabular} & \begin{tabular}[c]{@{}c@{}}Convolution\\[0.5ex] threshold\end{tabular} & \begin{tabular}[c]{@{}c@{}}Positive\\[0.5ex] detected \\[0.5ex] images\end{tabular} & \begin{tabular}[c]{@{}c@{}}Relative\\[0.5ex] CMEs \\[0.5ex] detected\\[0.5ex] RCD (\%)\end{tabular} & \begin{tabular}[c]{@{}c@{}}Absolute\\[0.5ex] CMEs \\[0.5ex] detected\\[0.5ex] ACD (\%)\end{tabular} & \begin{tabular}[c]{@{}c@{}}Extra\\[0.5ex] Detected\\[0.5ex] ED (\%)\end{tabular} & \begin{tabular}[c]{@{}c@{}}Reduced\\[0.5ex] telemetry\\[0.5ex] RT (\%)\end{tabular} & Remarks \\ \hline
\multirow{3}{*}{1} & \multirow{3}{*}{\begin{tabular}[c]{@{}c@{}}2010-11-10\\[1ex] 284\end{tabular}} & \multirow{3}{*}{27} & 0.4 & 239 & 92.59 & 8.8 & 89.53 & 84.15 & \multirow{3}{*}{Faint CME} \\
 &  &  & 0.5 & 58 & 70.37 & 6.69 & 67.24 & 20.42 &  \\
 &  &  & 0.6 & 21 & 44.44 & 0.7 & 42.85 & 7.39 &  \\ \hline
\multirow{3}{*}{2} & \multirow{3}{*}{\begin{tabular}[c]{@{}c@{}}2010-11-11\\[1ex] 284\end{tabular}} & \multirow{3}{*}{108} & 0.4 & 255 & 100 & 38.02 & 57.64 & 89.78 & \multirow{3}{*}{\begin{tabular}[c]{@{}c@{}}Medium\\[0.5ex] to bright\\[0.5ex] CMEs\end{tabular}} \\
 &  &  & 0.5 & 152 & 94.44 & 35.91 & 32.89 & 53.52 &  \\
 &  &  & 0.6 & 109 & 84.25 & 32.04 & 16.51 & 38.38 &  \\ \hline
\multirow{3}{*}{3} & \multirow{3}{*}{\begin{tabular}[c]{@{}c@{}}2010-11-12\\[1ex] 284\end{tabular}} & \multirow{3}{*}{56} & 0.4 & 213 & 94.64 & 18.66 & 75.11 & 75 & \multirow{3}{*}{\begin{tabular}[c]{@{}c@{}}Fast\\[0.5ex] faint\\[0.5ex] CMEs\end{tabular}} \\
 &  &  & 0.5 & 78 & 83.92 & 16.54 & 39.74 & 27.46 &  \\
 &  &  & 0.6 & 52 & 75 & 14.78 & 19.23 & 18.30 &  \\ \hline
\multirow{3}{*}{4} & \multirow{3}{*}{\begin{tabular}[c]{@{}c@{}}2011-08-02\\[1ex] 280\end{tabular}} & \multirow{3}{*}{47} & 0.4 & 227 & 87.23 & 14.64 & 81.93 & 81.07 & \multirow{3}{*}{\begin{tabular}[c]{@{}c@{}}Bright\\[0.5ex] and faint\\[0.5ex] CMEs\end{tabular}} \\
 &  &  & 0.5 & 63 & 65.95 & 11.07 & 50.79 & 22.5 &  \\
 &  &  & 0.6 & 34 & 61.70 & 10.35 & 14.70 & 12.14 &  \\ \hline
\multirow{3}{*}{5} & \multirow{3}{*}{\begin{tabular}[c]{@{}c@{}}2011-08-03\\[1ex] 280\end{tabular}} & \multirow{3}{*}{23} & 0.4 & 217 & 100 & 8.21 & 89.4 & 77.5 & \multirow{3}{*}{Bright CME} \\
 &  &  & 0.5 & 54 & 100 & 8.21 & 57.40 & 19.28 &  \\
 &  &  & 0.6 & 22 & 91.30 & 7.5 & 4.54 & 7.85 &  \\ \hline
\multirow{3}{*}{6} & \multirow{3}{*}{\begin{tabular}[c]{@{}c@{}}2011-08-04\\[1ex] 280\end{tabular}} & \multirow{3}{*}{49} & 0.4 & 219 & 97.95 & 17.14 & 78.08 & 78.21 & \multirow{3}{*}{Bright CME} \\
 &  &  & 0.5 & 53 & 61.22 & 1.07 & 43.39 & 18.92 &  \\
 &  &  & 0.6 & 26 & 40.81 & 7.14 & 23.07 & 9.28 &  \\ \hline
\multirow{3}{*}{7} & \multirow{3}{*}{\begin{tabular}[c]{@{}c@{}}2012-06-13\\[1ex] 2012-06-14\\[1ex] 2012-06-15\\[1ex] \\ 771\end{tabular}} & \multirow{3}{*}{77} & 0.4 & 612 & 100 & 9.98 & 87.41 & 79.37 & \multirow{3}{*}{\begin{tabular}[c]{@{}c@{}}Medium\\[0.5ex] and bright\\[0.5ex] CMEs\end{tabular}} \\
 &  &  & 0.5 & 193 & 93.5 & 9.33 & 62.69 & 25.03 &  \\
 &  &  & 0.6 & 90 & 80.51 & 8.04 & 31.11 & 11.67 &  \\ \hline
\end{tabular}
\end{sidewaystable}

Similar to high jitter we test the algorithm on several very faint, faint and bright CMEs. The results of table \ref{low jitter table} are
\begin{itemize}
\item All the CMEs identified manually are also identified by the automated algorithm.
\item Detection efficiency of images for a corresponding CME is lowest for faint CME. For very faint, detection efficiency is found to be $\approx$ 80 $\%$ , 14 $\%$ and 3 $\%$ for convolution thresholds of 0.4, 0.5 and 0.6 respectively. 
\item For faint event, detection efficiency is  $\approx$ 90 $\%$ , 75 $\%$ and 60 $\%$ for convolution thresholds of 0.4, 0.5 and 0.6 respectively.
\item For bright events, detection efficiency is  $\approx$ 100 $\%$ , 100 $\%$ and 90 $\%$ for convolution thresholds of 0.4, 0.5 and 0.6 respectively.
\item It might seem that convolution threshold of 0.4 is detecting CMEs with highest efficiency. However it should be noted that the data volume reduction is much less for the convolution threshold of 0.4. Thus it is a trade off between downloaded data and CME detection efficiency. 
\item The average downloaded data volume for convolution threshold of 0.4, 0.5 and 0.6 is $\approx$ 80 $\%$, 20 $\%$ and 10 $\%$ respectively. 
\item We find that except for very faint events, the threshold of 0.5 is good enough to capture the CMEs and at the same time is able to reduce data volume to about 20$\%$. Further reduction by a factor of 2 can be done by Rice compression. Thus total data volume will be reduced to 10 $\%$ in case of low jitter.
\end{itemize}

Thus total data volume can be reduced to 10 $\%$ and 17 $\%$ for low and high jitter, respectively, in the images.

\section{Application to Simulated Coronagraph Images for VELC FOV}
\label{S-Results simulated} 
    
\begin{figure}   
  \centerline{\hspace*{0.015\textwidth}
              \includegraphics[width=0.4\textwidth,clip=]{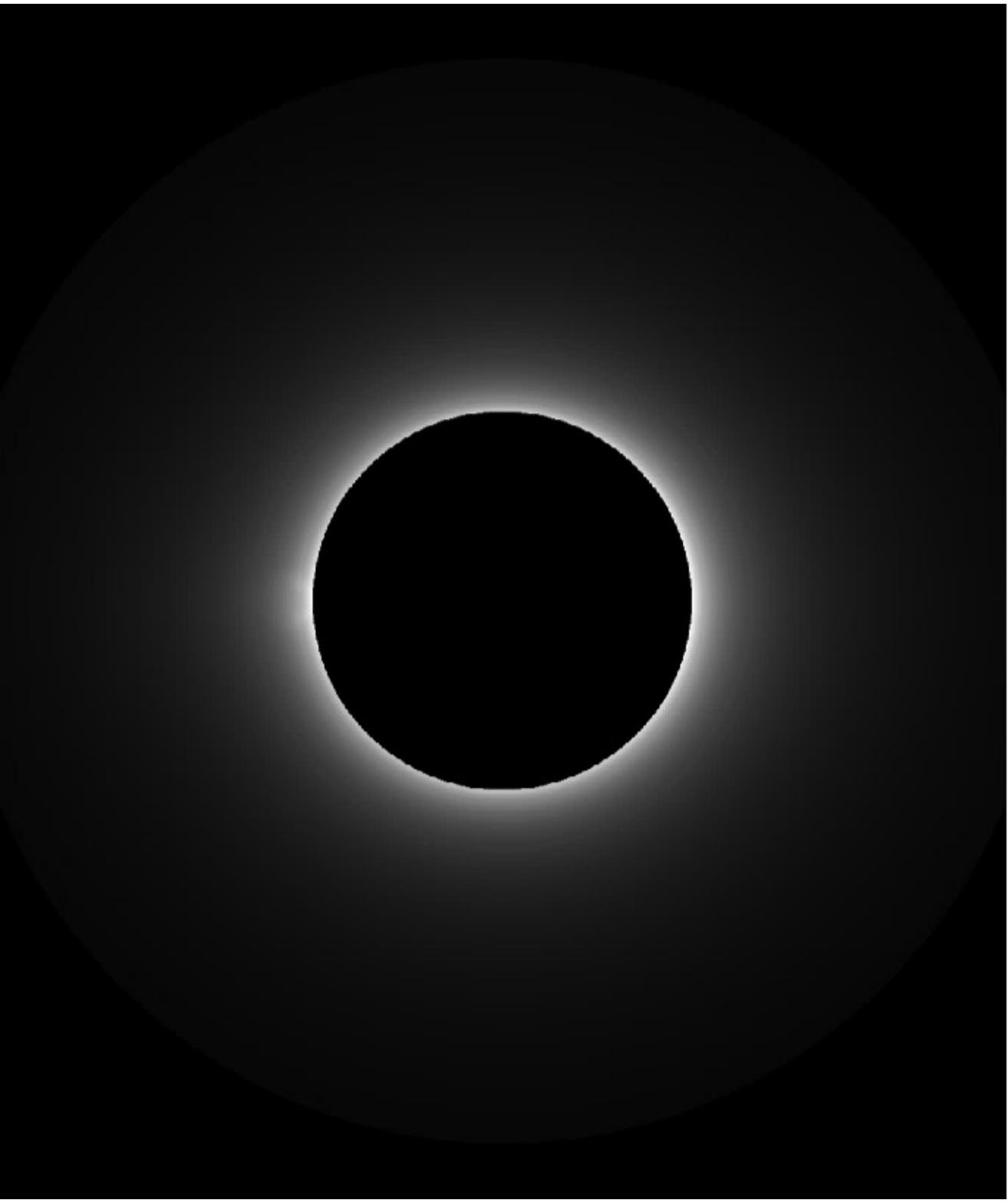}
              \hspace*{-0.0001\textwidth}
              \includegraphics[width=0.4\textwidth,clip=]{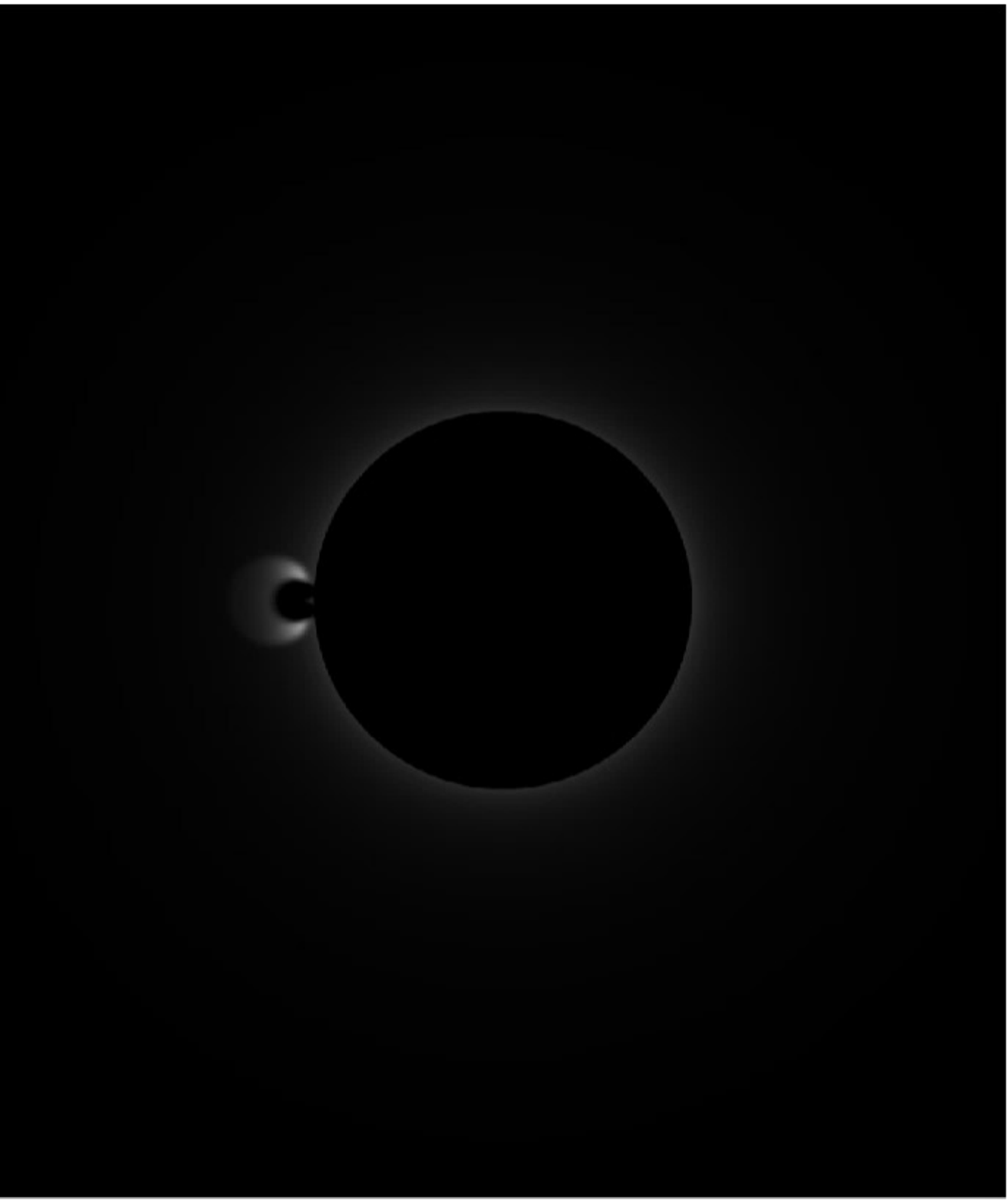}
              }
     \vspace{-0.08\textwidth}  
     \centerline{\Large     
      \hspace{0.09 \textwidth}  \color{white}{(a)}
      \hspace{0.32\textwidth}  \color{white}{(b)}
         \hfill}
     \vspace{0.015\textwidth}     
  \centerline{\hspace*{0.015\textwidth}
              \includegraphics[width=0.4\textwidth,clip=]{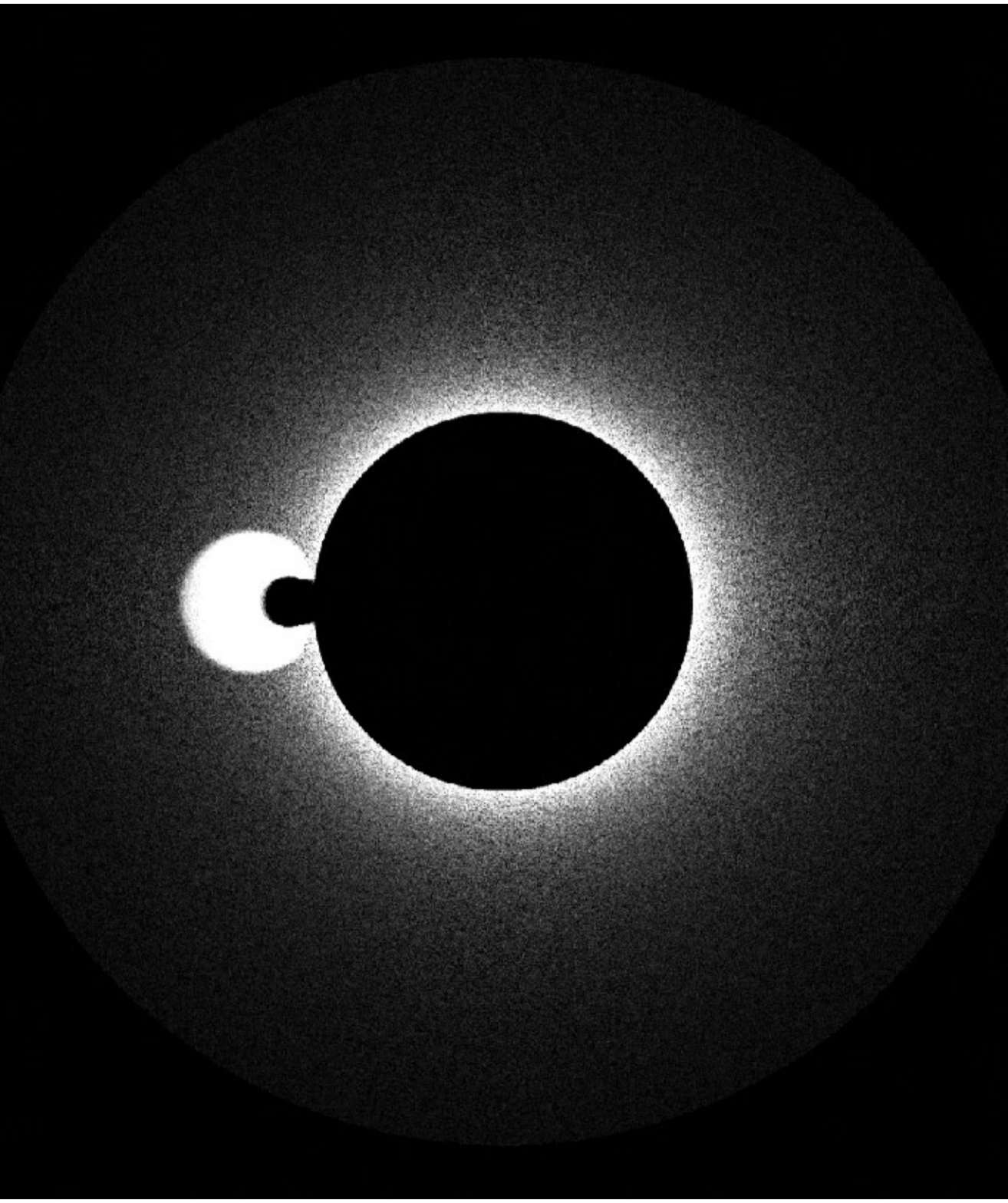}
              \hspace*{-0.0001\textwidth}
              \includegraphics[width=0.4\textwidth,clip=]{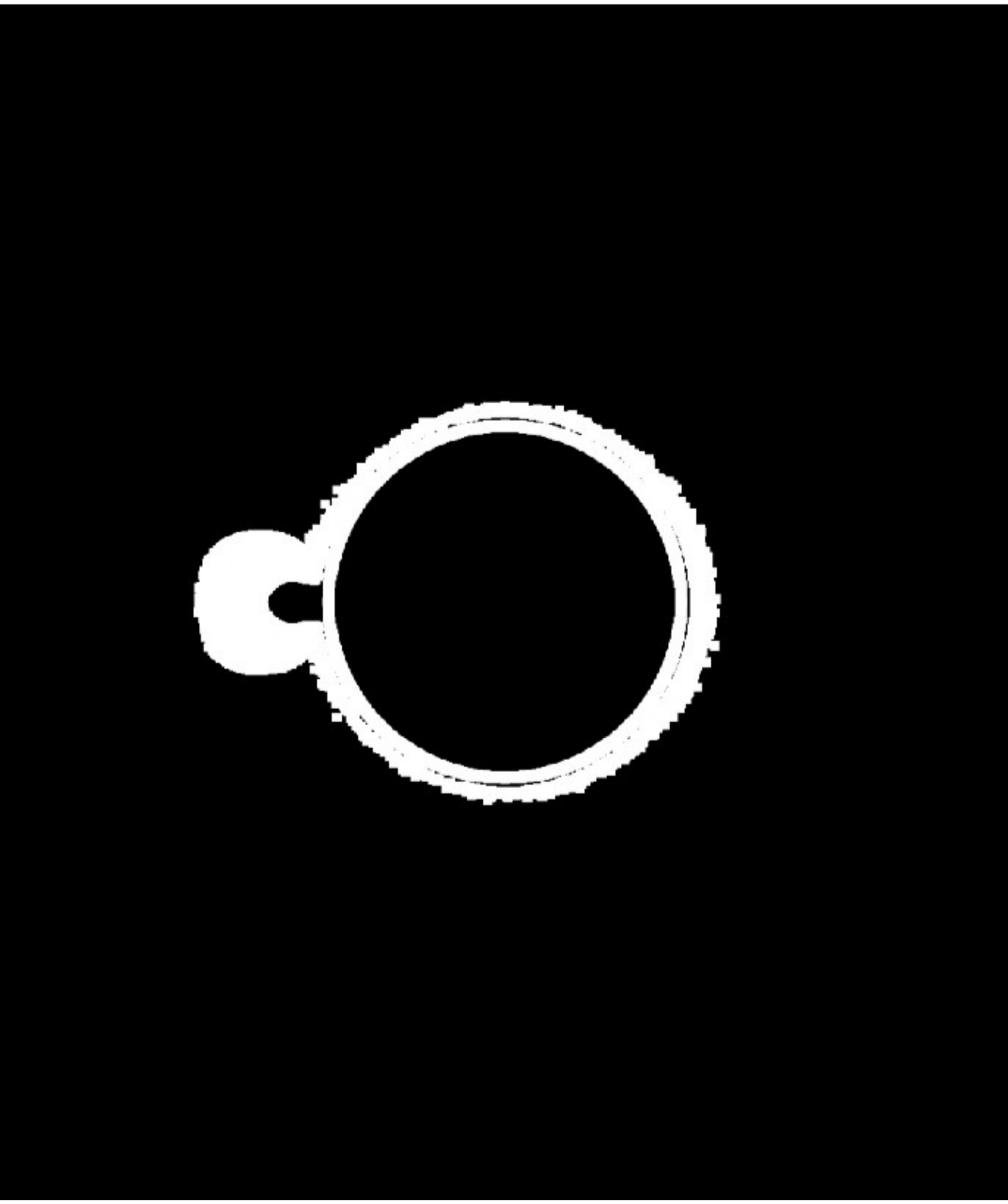}
              }
     \vspace{-0.08\textwidth}    
     \centerline{\Large     
      \hspace{0.09 \textwidth} \color{white}{(c)}
      \hspace{0.32\textwidth}  \color{white}{(d)}
         \hfill}
     \vspace{0.015\textwidth}    
  \centerline{\hspace*{0.015\textwidth}
              \includegraphics[width=0.4\textwidth,clip=]{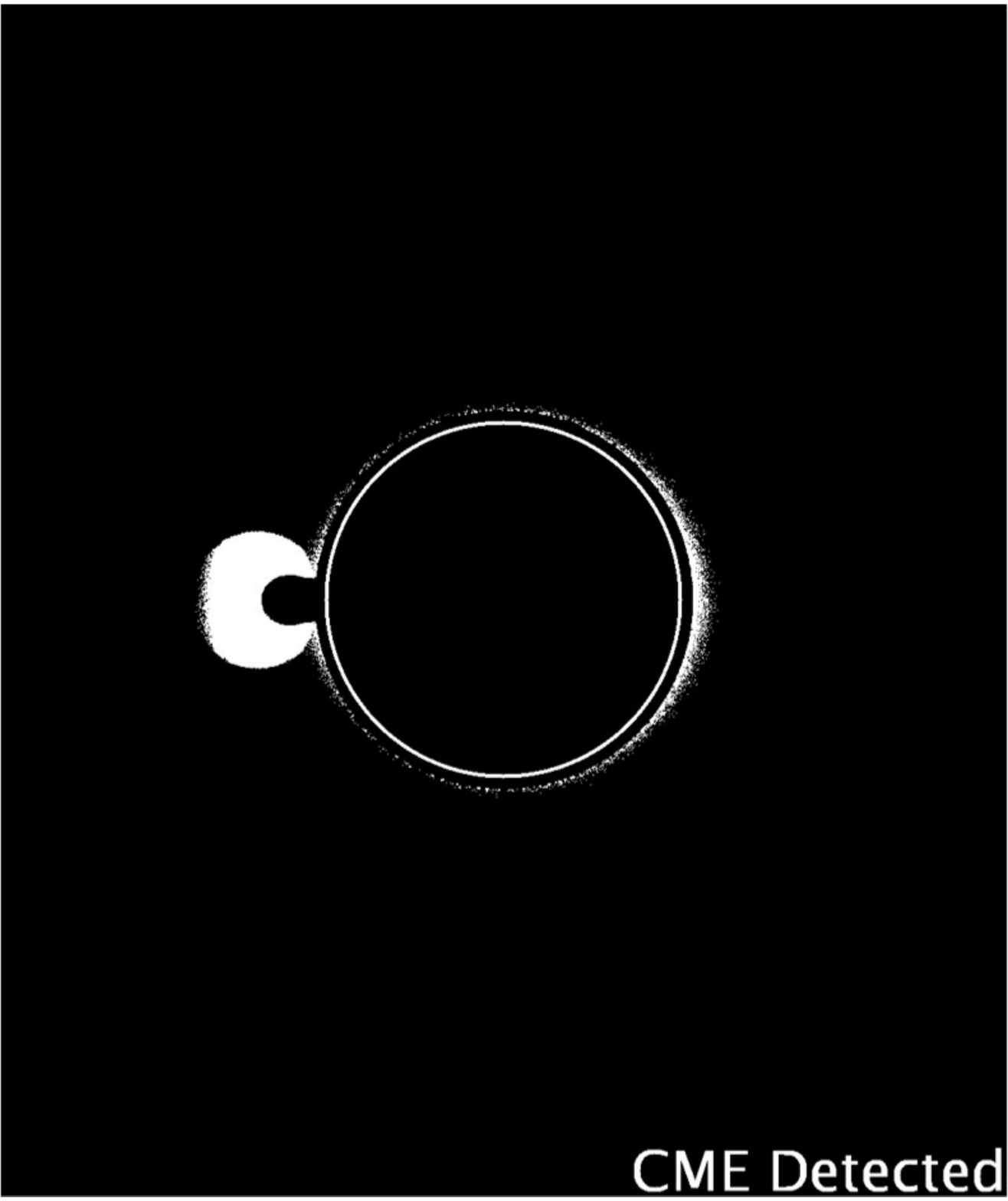}
              \hspace*{-0.0001\textwidth}
              \includegraphics[width=0.4\textwidth,clip=]{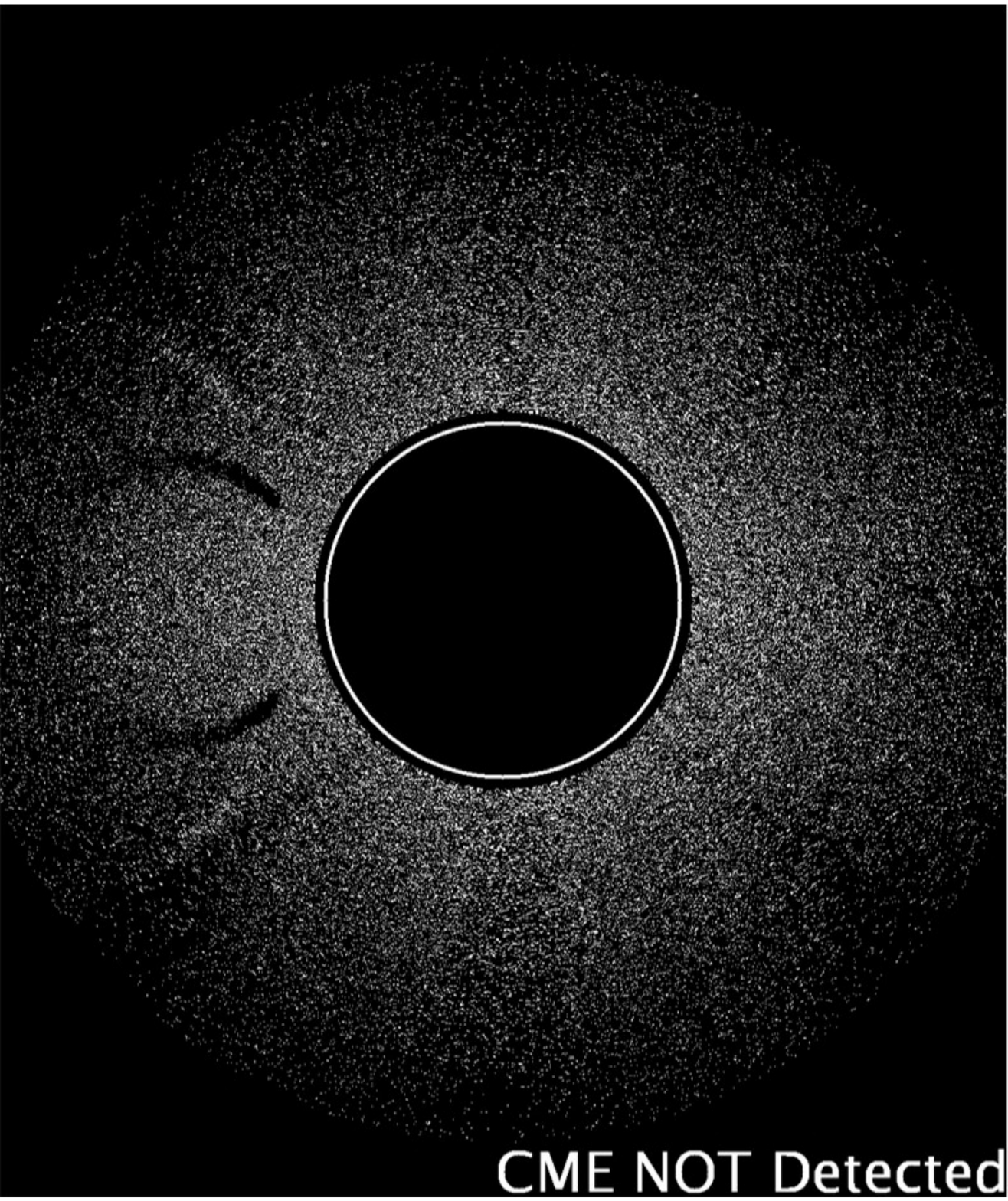}
              }
     \vspace{-0.08\textwidth}    
     \centerline{\Large      
      \hspace{0.09 \textwidth} \color{white}{(e)}
      \hspace{0.32\textwidth}  \color{white}{(f)}
         \hfill}
     \vspace{0.015\textwidth}    
              
\caption{(a) Synthetic coronal image after addition of ten frames and spatial bining of $2\times2$ pixels. (b) Difference image at an interval of 120 s (c) Intensity threshold image with factor 1. (d) Binary image convolved with kernel of size $10\times10$. (e) CME detected in the synthetic coronal image. (f) CME not detected in synthetic coronal image as CME almost merged with the background.
        }
  \label{Application}
  \end{figure}

We tested the algorithm on simulated CMEs images which consists of CMEs of varying brightness and average speed. Figure \ref{Application} illustrates an application of this algorithm. The Figure \ref{Application}a is obtained after average of ten images of 1 s equivalent exposure time and then re-binning it by $2\times2$ pixels. No feature can be seen in these images as the contribution of scattered intensity is dominant. The successive images are re-binned and then running difference of images 120 s apart is obtained as shown in Figure \ref{Application}b. The CME which wasn't visible in previous image due to high scattered brightness, is visible in this image. It is then converted to binary image by taking intensity threshold value equal to mean + factor$\times $(standard deviation)$ $ where factor is taken as 1, and is shown in Figure \ref{Application}c. This is then convolved with a kernel of size $10\times10$ and the resulting image is shown in Figure \ref{Application}d. The value of convolution threshold in this case is taken as 0.8 and the maximum of the convolved image is compared with this value. In case Figure \ref{Application}e, when the maximum of the convolved image is more than the threshold value, CME has been detected. On the other hand, when CME has almost crossed VELC FOV and merged with the background, the detection is stopped as shown in Figure \ref{Application}f.

Since, the running difference is performed on images of 10 s equivalent exposure time images, so 6 images will be generated in 1 minute. Considering 24 hours of operation of the instrument a total of 8640 images ($24\times60\times6$) will be generated each day. The results presented in table \ref{simulated table} are based on the total number of images as 8640. The detection efficiency mentioned here are the fraction of CMEs images which are detected by the algorithm which is the relative CMEs detection like previous cases. After the analysis we find that

\begin{itemize}
\item CMEs detected by the visual inspection are also detected by the automated detection algorithm.
\item No false detection has been noticed with the chosen parameters for simulated CMEs. This may be due to absence of other features like streamers which will be there in real data and are sometimes responsible for false detection. 
\item Faint CMEs, both normal and narrow width, were detected with efficiency less than 70\%.
\item Slow CME (speed $\approx$ 100 km { s}$^{-1}$) was detected with maximum detection efficiency of $\approx$69\% when the difference images were taken at an interval of 120 s. The same jumped to 95\% when the interval was increased to 300 s.
\item Detection efficiency of very bright CMEs is more than 90 $\%$ with factor of 1 and taking convolution threshold of 0.8 for kernel size of $10\times10$. This efficiency of detection of bright CMEs is seen for both fast CMEs (speed $\approx$ 2000 km { s}$^{-1}$) and average speed CMEs (speed $\approx$ 400 km { s}$^{-1}$).
\item The running difference interval of 120 s can be said to be ideal for fast CMEs. However, it is seen that the detection efficiency of average speed and slow CMEs increases with increase in this interval up to 300 s.
\item It can be seen that increasing the value of factor from 1 to 2.5 results in the decrease in CMEs detection efficiency in all the cases.
\item   { Halo CMEs are detected with less efficiency except for factor of 1. Since these CMEs cover most part of images, the difference images may not satisfy higher threshold values for more images.}
\item For this analysis it has been assumed that VELC will produce 1 image every 10 s, so a total of 8640 images will be generated in 24 hours of its operation. So, even for the slow CMEs, considering highest detection efficiency of 95\%, data volume reduces to $\approx$15\% which is also close to 1300 images, assuming it to be the only CME in that 24 hours. For other cases for simulated CMEs, it will be even less.
\end{itemize}

\begin{table}[]
\centering
\resizebox{\columnwidth}{!}{%
\setlength{\tabcolsep}{4pt}
\caption{Application of CMEs detection algorithm to simulated CMEs images at convolution threshold of 0.8 and Kernel size of $10\times10$.}
\label{simulated table}
\begin{tabular}{cccccccccccccccc}
\hline
\multirow{3}{*}{S.No.} & \multirow{3}{*}{CME} & \multirow{3}{*}{\begin{tabular}[c]{@{}c@{}}Total\\[1ex] number \\[1ex] of CME\\[1ex] images\end{tabular}} & \multirow{3}{*}{\begin{tabular}[c]{@{}c@{}}Difference\\[1ex] interval\\[1ex] (s)\end{tabular}} & 

\multicolumn{4}{c}{\begin{tabular}[c]{@{}c@{}}Relative CMEs detection \\[0.5ex] RCD (\%) \end{tabular}} & \multicolumn{4}{c}{\begin{tabular}[c]{@{}c@{}}Reduced Telemetry \\[0.5ex] RT (\%) \end{tabular}}\\[1.5ex] 
 &  &  &  & \multicolumn{4}{c}{Factor} &   \multicolumn{4}{c}{Factor} \\
 &  &  &  & 1 &1.5 & 2 & 2.5 &  1 & 1.5 & 2 & 2.5 \\ \hline

\multirow{2}{*}{1} & \multirow{2}{*}{CME 1} & \multirow{2}{*}{1385} & 120 & 69.1 & 63.9 & 61.7 & 55.7 & 11.0 & 10.2 & 9.8 & 8.9 \\ 
 &  &  & 300 & 95.3 & 89.7 & 81.6 & 77.6 & 15.2 & 14.3 & 13.1 & 12.4\\ \hline
2 & CME 2 & 98 & 120 & 68.3 & 63.2 & 50 & 48.9 & 0.77 & 0.71 & 0.56 & 0.55\\ \hline
\multirow{2}{*}{3} & \multirow{2}{*}{CME 3} & \multirow{2}{*}{569} & 120 & 62.9 & 61.3 & 56.5 & 53.1 & 4.1 & 4.0 & 3.7 & 3.4\\ 
 &  &  & 300 & 71.8 & 67.8 & 62.2 & 58.6 & 4.7 & 4.4 & 4.1 & 3.8\\ \hline
4 & CME 4 & 108 & 120 & 62.0 & 55.5 & 45.3 & 44.4 & 0.77 & 0.69 & 0.56 & 0.55\\ \hline
5 & CME 5 & 192 & 120 & 96.8 & 90.1 & 80.7 & 74.4 & 2.15 & 2.00 & 1.79 & 1.65\\ \hline
6 & CME 6 & 131 & 120 & 86.7 & 73.5 & 66.1 & 61.9 & 1.31 & 1.11 & 1.00 & 0.93\\ \hline
7 & CME 7 & 143 & 120 & 93.3 & 87.6 & 79.3 & 73.5 & 1.54 & 1.44 & 1.31 & 1.21\\ \hline
{ 8} & { CME 8} & { 101} & { 120} & { 88.1} & { 34.6} & { 24.7} & { 20.7} & { 1.03} & { 0.41} & { 0.28} & { 0.24}\\ \hline
{ 9} & { CME 9} & { 109} & { 120} & { 78.9} & { 48.6} & { 31.2} & { 26.6} & { 0.99} & { 0.61} & { 0.39} & { 0.33}\\ \hline
\end{tabular}
}
\end{table}

\section{Summary and Conclusions}
      \label{S-Conclusion} 
      
In this work, a novel onboard CME detection algorithm is presented which can detect CMEs in coronagraph images. This will be one of the first algorithms to be implemented in onboard electronics. The detection algorithm based on intensity thresholding and area thresholding is kept simple so that it can be implemented in onboard electronics. The performance of this algorithm is tested on  existing space and ground based coronagraph images, and synthetic coronagraph images for different choices of free parameters. The analysis shows that this algorithm is capable of detecting CMEs with detection efficiency as high as 100$\%$ and reducing the telemetry to 15\% and better at the same time. As VELC is capable of taking images of corona with high spatial ($\approx$ 2.51 arcsec pixel$^{-1}$) and temporal resolution (1 s), more than 1TB of data will be generated in a day from the continuum channel alone. Such high data volume can be reduced by 85\% and better by implementing this onboard automated CME detection algorithm. Moreover, such algorithm can enable us to take even higher resolution images with limited telemetry in the future.

We have applied the algorithm on existing data of STEREO COR-1A and tested it on the images with low and high jitters. The reduced telemetry for high jitter cases vary from $\approx$10\% to $\approx$90\% while for low jitter cases it varies from  $\approx$10\% to $\approx$80\%. We applied this algorithm on few images taken by K-Cor that has same FOV as VELC. The algorithm could not be applied to most of the  dataset of K-Cor because the images are affected by atmospheric contamination which resulted in false detection as high as $\approx$90\%. The noisy patterns in these images were too dominant to satisfy the intensity and convolution thresholds.

Since, no other space-based coronagraphs has FOV similar to VELC, as an alternative test we created synthetic coronal images. We modified the Huslt model of solar corona such that the gradual variation of intensity from equator to poles can be incorporated in a single equation and created synthetic corona images for VELC FOV. CMEs of different intensities with respect to background corona, speeds and shapes were simulated on these simulated images. The algorithm was then tested on these simulated images. 

We have illustrated that this algorithm can be easily implemented for CMEs detection as applied to observed data or simulated data by varying the free parameters in the algorithm.  
The value of factor was varied as 1, 1.5 and 2 with the convolution threshold as 0.6 for kernel size of $10\times10$ pixels for its application to K-Cor images. For the application on COR-1A data we varied the value of convolution threshold as 0.4, 0.5 and 0.6 keeping the value of factor as 1.5 and size of kernel as $8\times8$ pixels. While applying the algorithm on simulated images, we varied the value of factor as 1, 1.5, 2 and 2.5 with the convolution threshold as 0.8 for kernel size of $10\times10$ pixels. It should be noted that the algorithm does not count the number of CMEs occurring in a day{ , nor it can differentiate between narrow, normal and halo CMEs}. Its sole purpose is to detect the images containing single or multiple CMEs and to reduce the data volume. Though we found that multiple CMEs can be detected by the algorithm once the thresholds are satisfied.

It has been found that there can be some loss of data of interest as the CMEs after reaching the end of FOV of the coronagraph fails to meet threshold criteria and hence results in data loss. No combination of free parameters can always result in 100 $\%$ detection of CMEs in images. Very faint CMEs had detection efficiency of even less than 10$\%$ whereas the bright CMEs were detected with efficiency of more than 90$\%$ and sometimes even 100$\%$. This also implies that the detection efficiency can vary with the solar cycle as bright CMEs are more prominent during solar maxima than minima. It was also found that the CMEs detection efficiency increases when the interval at which the difference image is taken is increased for slow and average speed CMEs. Moreover, the low detection efficiency in COR-1A images with faint CMEs is due to jitters in images which restricts the lower limit of parameter values to avoid detection of noise. Therefore, a compromise has to be made between the percentage detection and telemetry depending on the science requirements over different phases of solar cycle. The values of the free parameters are therefore kept tunable so that they can be changed post launch depending on the period of operation and noise levels recorded in the instrument. In case of VELC, it has been proposed to store few images even after the algorithm stops detection of CMEs. Though this will increase some volume of data, but can reduce the loss of CMEs images which are missed in the onboard detection.  

We should also point out that although  the algorithm has been designed to detect  the presence of CMEs in coronagraph images, it may also show false triggers due to surging material in coronal loops which may enter VELC FOV, cosmic ray hits, passing planet in FOV and various other sources which may satisfy the threshold conditions. The intensity threshold though is accompanied by area threshold to avoid most of these false detection, sometimes we may get images which may not contain CMEs but other bright sources as mentioned above. However, chances of such detection are very low, but an analysis can be done using the existing space-based coronagraph data with such bright sources (except for coronal loops material) in the future. The corona model can be further improved by introducing streamer like structures which varies over time. The performance of this algorithm can be put to such test further. Moreover, the synthetic corona model can be further improved to take in to account the temporal variation of the coronal intensity over different phases of solar cycle.

The overall analysis shows reduction in the data volume for on board storage and subsequent download through limited telemetry  which is the main objective of this algorithm. With ADITYA launching in 2020, this algorithm will be the first onboard automated CME detection algorithm.

\begin{acks}
 We thank the ADITYA project team.  ADITYA L1 is project of ISRO. We would also like to thank the Mauna Loa Solar Observatory, operated by the High Altitude Observatory, as part of the National Center for Atmospheric Research (NCAR). NCAR is supported by the National Science Foundation. We also acknowledge SECCHI/STEREO consortium for providing data. STEREO is a project of NASA. The SECCHI data used here were produced by an international consortium of the Naval Research Laboratory (USA), Lockheed Martin Solar and Astrophysics Lab (USA), NASA Goddard Space Flight Center (USA), Rutherford Appleton Laboratory (UK), University of Birmingham (UK), Max-Planck-Institut for Solar System Research (Germany), Centre Spatiale de Li$\grave{e}$ge (Belgium), Institut d'Optique Th$\acute{e}$orique et Appliqu$\acute{e}$e (France), Institut d'Astrophysique Spatiale (France).
\end{acks}

\vspace{0.25cm}
\footnotesize
{ Disclosure of Potential Conflicts of Interest}
The authors declare that they have no conflicts of interest.



\end{article} 
\end{document}